\pdfoutput=1
\documentclass[preprint,12pt]{elsarticle}

\usepackage{amssymb}
\usepackage{soul}
\usepackage{natbib}
\usepackage[table]{xcolor}
\usepackage{amsmath}
\usepackage{booktabs}
\usepackage{multirow}
\usepackage{tabularx}
\usepackage{longtable}
\usepackage{url}
\usepackage{footnote}
\usepackage{caption}
\usepackage{subcaption}
\usepackage[hidelinks]{hyperref}
\usepackage{graphicx}
\usepackage[nolist]{acronym}
\usepackage[table]{xcolor}
\usepackage[T1]{fontenc}
\usepackage[utf8]{inputenc}
\usepackage{rotating}
\usepackage{colortbl}
\DeclareCaptionFont{black}{\color{black}}

\usepackage{soul}
\usepackage{threeparttable}

\usepackage{amssymb}
\usepackage{pifont}

\newcommand{\xmark}{\ding{55}}

\newacro{ASR}[ASR]{automatic speech recognition}
\newacro{CTS}[CTS]{conversational telephone speech}
\newacro{BIC}[BIC]{Bayesian information criterion}
\newacro{DER}[DER]{Diarization Error Rate}
\newacro{MFCC}[MFCC]{Mel-frequency cepstral coefficients}
\newacro{AHC}[AHC]{agglomerative hierarchical clustering}
\newacro{PIT}[PIT]{permutation invariant training}
\newacro{SSGD}[SSGD]{speech separation guided diarization}
\newacro{TDNN}[TDNN]{time-delay neural network}
\newacro{VAD}[VAD]{voice activity detection}
\newacro{SRE}[SRE]{Speaker Recongition Evaluation}
\newacro{DNN}[DNN]{deep neural network} 

\journal{Computer Speech \& Language}
\begin{document}

\begin{frontmatter}

\title{An Experimental Review of Speaker Diarization methods with application to \textcolor{black}{Two-Speaker} Conversational Telephone Speech recordings}

\affiliation[1]{organization={Università Politecnica delle Marche},
            city={Ancona},
            country={Italy}}
            
\affiliation[2]{organization={PerVoice S.p.A.},
    city={Trento},
    country={Italy}}

\affiliation[3]{organization={Fondazione Bruno Kessler},
    city={Trento},
    country={Italy}}

\author[1]{Luca Serafini}
\ead{l.serafini@univpm.it}

\author[1]{Samuele Cornell}
\ead{s.cornell@pm.univpm.it}

\author[1]{Giovanni Morrone}
\ead{g.morrone@univpm.it}

\author[2]{Enrico Zovato}
\ead{enrico.zovato@pervoice.it}

\author[3]{Alessio Brutti}
\ead{brutti@fbk.eu}

\author[1]{Stefano Squartini}
\ead{s.squartini@univpm.it}

\begin{abstract}
We performed an experimental review of current diarization systems for the conversational telephone speech (CTS) domain.
In detail, we considered a total of eight different algorithms belonging to clustering-based, end-to-end neural diarization (EEND), and speech separation guided diarization (SSGD) paradigms. We studied the inference-time computational requirements and diarization accuracy on four CTS datasets with different characteristics and languages.
We found that, among all methods considered, EEND-vector clustering (EEND-VC) offers the best trade-off in terms of computing requirements and performance. 
More in general, EEND models have been found to be lighter and faster in inference compared to clustering-based methods. However, they also require a large amount of diarization-oriented annotated data. In particular EEND-VC performance in our experiments degraded when the dataset size was reduced, whereas self-attentive EEND (SA-EEND) was less affected. 
We also found that SA-EEND gives less consistent results among all the datasets compared to EEND-VC, with its performance degrading on long conversations with high speech sparsity. 
Clustering-based diarization systems, and in particular VBx, instead have more consistent performance compared to SA-EEND but are outperformed by EEND-VC. The gap with respect to this latter is reduced when overlap-aware clustering methods are considered. SSGD is the most computationally demanding method, but it could be convenient if speech recognition has to be performed. Its performance is close to SA-EEND but degrades significantly when the training and inference data characteristics are less matched.

\end{abstract}


\end{frontmatter}

\section{Introduction}\label{sec:intro}

Speaker diarization, also often referred to as simply diarization, aims at partitioning an audio recording into temporal segments denoting the boundaries of each speaker's utterances. In other words, it addresses the problem of ``who spoke when?'', without any a-priori knowledge of the speakers' identities. 
This discipline represents an important research field in the speech processing area. Its main applications involve \ac{ASR}, speaker indexing, speaker recognition, real-time captioning, and audio analysis to name a few. 
Speaker diarization constitutes an important and often essential pre-processing step in most of these application scenarios: e.g., accurate diarization can be used effectively to drive multi-channel blind source separation algorithms to separate concurrent speakers for distant speech recognition \cite{boeddeker2018front, kanda2019guided} or speaker adaptation for ASR~\cite{saon2013speaker, miao2015speaker, wang2017unsupervised, sari2020unsupervised}.
It is also a very challenging task as speaker identity is arguably not easy to estimate especially when many speakers are present and they often overlap. In such scenarios, even humans struggle and have to resort to linguistic information from neighboring utterances or long term information to resolve speaker identities' ambiguities on short segments. 

As such, it can be argued that the two-speaker conversation case is the simplest scenario when all the other characteristics that can impact diarization performance are equal: e.g., audio quality due to the registration setup, environmental noise, duration and sequence of speakers' turns, as well as the style and spontaneity of speech.
On the other hand, the two-speaker scenario constitutes a particularly important use-case as it is a very common scenario and has a great commercial interest. 
In fact, it represents the most ordinary situation in normal conversations that take place via telephone or remotely via hands-free tele-conferencing, but it also appears in many other contexts, such as in doctor-patient meetings, interviews, air traffic control conversations, et cetera.

In light of the tremendous progress made in recent years, we present in this work an experimental review focused on \ac{CTS}, and compare current state-of-the-art methods based on clustering, separation, and end-to-end diarization on real-world \ac{CTS} datasets with different languages. 
The main motivation is that recent research trends, such as the invention of fully end-to-end diarization methods \cite{fujita2019end-blstm, fujita2019end-self, fujita2020end, kinoshita2021integrating, kinoshita2021advances, kinoshita2022utterance, kinoshita2022tight}, seems to suggest that the technology is mature enough to attain highly accurate and reliable diarization in the \ac{CTS} scenario, whereas other scenarios, such as distant-talk meetings with several speakers, still pose several challenges \cite{watanabe2020chime}. Our hope is that this work will be useful to researchers and practitioners alike, as we investigate computational requirements and the performance of eight diarization algorithms on multilingual data, all belonging to the \ac{CTS} domain but with diverse characteristics, such as overlapped speech ratio, average length of the conversation, and speech sparsity.

The remainder of the paper is organized as follows: we present in Section \ref{sec:history} and \ref{sec:motivation} a brief overview of speaker diarization methods and a comparison of this review work with previous ones.
Section \ref{sec:section_2} describes the various systems that have been considered in our experimental analysis, along with their main characteristics. In Section \ref{sec:section_3} we outline the experimental setup and describe the systems implementation and hyper-parameters employed in our experiments, as well as the diarization metrics and the \ac{CTS} datasets used for training and testing.
In Section \ref{sec:results_discussion} we analyze the results.
Finally, in Section \ref{sec:conclusion} we draw conclusions and discuss possible future research directions.

\subsection{A Brief History of Speaker Diarization}\label{sec:history}
The first works on speaker diarization can be traced back to the 1990s \cite{gish1991segregation, siu1992unsupervised, jain1996recognition, chen1998speaker, liu1999fast}. 
These early works focused on applications such as radio broadcast news and communications, with the main goal of improving ASR performance. 
As such, some of these works relied directly on ASR outputs \cite{jain1996recognition, liu1999fast}, e.g., by performing a two-pass decoding, where the first pass was used for detecting speakers' changes.
Also focused on ASR, \cite{siu1992unsupervised} can be considered one of the very first works on \ac{SSGD} as the authors performed diarization using blind speech separation.
In these early years, the foundations for the clustering-based diarization paradigm were laid \cite{gish1991segregation, chen1998speaker}. 
These works formulated diarization as a clustering problem. After \ac{VAD}, speaker discriminative features are extracted from each speech segment. These latter are then clustered together in order to assign each of the features and their corresponding sub-segments to each speaker and thus identify the number of speakers in the recording and obtain a segmentation for each speaker. 
The features used in such early studies were largely ``hand-crafted", with \ac{MFCC} being a popular choice together with linear frequency cepstral coefficients (LFCC), perceptual linear predictive (PLP) \cite{hermansky1990perceptual}, and linear predictive coding (LPC) \cite{o1988linear}. 
Often a combination of more than one feature type \cite{yamaguchi2005spectral} was used in order to increase the system robustness by capturing various speech characteristics possibly at different time scales.
Regarding clustering, these early works focused on \ac{AHC} performed based on a distance measure between each feature vector belonging to a different segment. 
How to define such distance and what properties the feature vectors should have in order to have a well-defined distance from a speaker identity perspective is still an active research direction.
The use of \ac{BIC}~\cite{chen1998speaker} to define such distance and perform \ac{AHC} was found to be particularly effective at the time and remained a popular choice in subsequent years together with the generalized likelihood ratio (GLR) \cite{gish1991segregation}.

The first decade of the new millennium saw tremendous improvement and growth in the diarization field and research directions also included more challenging scenarios, such as meetings captured by far-field microphone arrays. 
For example, the still very relevant and popular AMI \cite{carletta2005ami} dataset was collected in 2005.
In this period the clustering-based diarization paradigm was consolidated and became the de-facto standard approach. 

One of the major improvements over previous techniques was due to the fact that researchers understood the necessity to move away from fully hand-crafted features and devise data-driven methods to obtain more robust and higher-level speaker-id discriminative features. 
One of the main challenges in obtaining good speaker identity features lies in the fact that these should be robust to intra-speaker variability (e.g., due to changes in intonation, background noise, etc.) while being enough discriminative to allow for differentiating between distinct speakers (i.e., inter-speaker discriminability).
Diarization systems started to incorporate machine learning models such as Gaussian mixture models (GMM). 
A key work was the one of Reynolds et al. \cite{reynolds2000speaker} which introduced the speaker-independent GMM-Universal Background Model (GMM-UBM) for speaker verification. 
In this work, each vector of features is derived in a data-driven fashion from a GMM (i.e., it consists of a weighted sum of a finite number of multi-dimensional Gaussian components).
Since in most applications there is often not enough data to train from scratch such a GMM model without overfitting, these feature vectors are obtained by adapting some parameters of a pre-trained GMM model: e.g., the means of the Gaussian components which, when concatenated are denoted as a GMM ``supervector".
The GMM-UBM model is trained on a large amount of data with a large speaker variability via a maximum a posteriori (MAP) criterion.
The use of a GMM-UBM remained a de-facto standard for speaker verification and diarization till the advent of \ac{DNN}-based techniques. 

One of the problems of such an approach is that it has the drawback of suffering from the aforementioned intra-speaker variability.
A number of important works \cite{kenny2007speaker, castaldo2008stream} tried to address this problem. For example, joint factored analysis (JFA) \cite{kenny2007speaker} and Eigenvoice priors \cite{castaldo2008stream} alleviate this issue by exploiting lower dimensional factorizations of the GMM supervectors to obtain more robust speaker-dependent supervectors which have lower intra-speaker variability. 
For example, in JFA, it is assumed that the supervector covariance matrix can be decomposed into a channel space and speaker space, where the channel space is responsible for the inter-speaker variations. This leads to the fact that each speaker and channel-dependent supervector can be decomposed into a sum of components that belongs to these two sub-spaces. As such, the two components can be disentangled and a more robust representation is obtained. 

However, Dehak et al. \cite{dehak2010front} found that the JFA channel factors also contain speaker-related information and instead propose to define just one total variability space which contains the channel and speaker factor of variations simultaneously.
They then propose to find, for each speaker utterance, through Baum-Welch statistics, a projection for this total variability space which is representative of the speaker's identity. This projection vector is called i-vector and was found to be a very effective speaker-id discriminative feature.
To compensate for channel-dependent variability, \cite{dehak2010front} proposes to use linear discriminant analysis (LDA) and within-class covariance normalization (WCCN) \cite{hatch2006within} where the i-vectors distances are defined with cosine similarity or with a support vector machine (SVM), respectively.
In \cite{matvejka2011full} it was shown that by using probabilistic LDA (PLDA) the performance can be improved further. 
This coupling of i-vectors and PLDA has been proven extremely effective and it has been a popular technique for speaker verification and diarization till the advent of deep learning based approaches.

Starting from 2014, the studies and refinements in the deep learning area, together with the increasing availability of annotated transcriptions and data, made it possible to exploit DNNs in place of GMMs for obtaining speaker-discriminative features. 
One of the first works in this direction is the one of Variani et al.~\cite{variani2014deep}, in which DNN-based features (so-called d-vectors) were shown to be able to outperform i-vectors, the state-of-the-art approach of the time, especially in noisy conditions. 
As it happens with the GMM-UBM, in \cite{variani2014deep} the DNN was trained on large corpora with a vast number of speakers and different acoustic conditions. 
The DNN is trained on fixed-length segments to try to classify the correct speaker (multi-class classification problem) among all the ones in the training set. 
In inference then the output of the last hidden layer is the d-vector and can be used as a speaker-discriminative feature for speaker verification or diarization. 
Follow-up works \cite{bai2021speaker} continued to improve this new paradigm based on DNNs used to extract speaker-identity discriminative features. 
In particular, a great improvement was due to the invention of the x-vector extractor \cite{snyder2018x} which employs a \ac{TDNN} and a statistical pooling layer to obtain a low-dimensional speaker-id representation. 
This trend of designing better DNN architectures to improve speaker-id discriminative features is continuing today. 
Some recent improvements include the use of ResNet-based designs \cite{landini2021analysis}, TitaNet~\cite{koluguri2022titanet}, ECAPA-TDNN~\cite{desplanques2020ecapa} and the use of self-supervised learning pre-trained models such as WavLM~\cite{chen2022wavlm}.
Other works \cite{bredin2017tristounet, li2018angular, chung2020defence} instead have focused on the loss function and training strategy to use to train such a DNN speaker-id feature extractor. For example, \cite{bredin2017tristounet, chung2020defence} proposed to use metric learning approaches, whereas \cite{li2018angular} the use of angular-softmax loss to improve performance. 
This is also a very active research direction. 
In all these works, clustering continued to be, as said, the main approach with DNN-based speaker-id representation used often in conjunction with PLDA to reduce dimensionality and intra-speaker variability before clustering, as done previously with i-vectors.

Indeed many advances also regarded other components of the diarization pipeline, such as the clustering step or the post-processing step. 

\textcolor{black}{For example, spectral clustering based on the Ng-Jordan-Weiss (NJW) algorithm \cite{ng2001spectral}, used in early works such as \cite{ning2006spectral}, was then replaced by spectral clustering with maximum eigengap.
Indeed, Wang et al. \cite{wang2018speaker} improved over the at the time state-of-the-art PLDA followed by AHC approach, by using this type of clustering procedure to aggregate DNN-extracted speaker embeddings, i.e., d-vectors.
Afterward, Park et. al.~\cite{park2019auto} also added an auto-tuning strategy to overcome the need for a development set for the hyper-parameter tuning.}

Another notable work, this time regarding post-processing, is variational Bayesian (VB) resegmentation, initially proposed for an i-vector-based system \cite{diez2018speaker} and later adapted to an x-vector-based system \cite{landini2020bayesian}. 
This latter approach called VBx has been proven extremely effective on a wide number of datasets \cite{landini2022simulated} and challenges \cite{ryant2019second}. 
Among the post-processing works it is worth mentioning ensembling or fusion methods that allow combining the output of multiple heterogeneous diarization systems in order to improve the performance.
Such works include diarization output voting error reduction (DOVER) \cite{stolcke2019dover} and DOVER-Lap, a recent improvement over DOVER that allows handling also overlapped speech regions \cite{raj2021dover, raj2021reformulating}.

A number of recent works also explored how to improve clustering-based methods with deep learning based techniques in order to let them deal also with overlapped speech. For example, Bullock et al.~\cite{bullock2020overlap} proposed to use an overlap detector to mask the speaker posterior matrix in the VBx method.
Raj et al.~\cite{raj2021multi} instead devised a way to handle speaker-id discriminative features from overlapped speech regions during the clustering step.

Due to the greater availability of annotated data and possibilities opened up by deep learning recently a new line of research arose, involving other alternative approaches to diarization that try a tighter integration with DNNs. 
Among such works are region-proposal networks diarization (RPND) \cite{huang2020speaker}, unbounded interleaved-state recurrent neural network (UIS-RNN) \cite{zhang2019fully}\textcolor{black}{, discriminative neural clustering (DNC) \cite{li2021discriminative}}, deep learning SSGD-based approaches \cite{fang2021deep, morrone2022low-latency} and target-speaker VAD (TS-VAD) \cite{medennikov2020target}.
Some of these approaches and in particular SSGD and TS-VAD have proven to be particularly effective: for example, the system winning the recent third DIHARD Challenge \cite{ryant2020third}, was based on a combination of SSGD and TS-VAD based approaches \cite{wang2021ustc}.

\textcolor{black}{Indeed TS-VAD has been, in the recent three years, the de-facto best approach in terms of raw diarization performance; moreover, all the top ranking submissions in DIHARD-III \cite{ryant2020third}, VoxSRC-21/22 \cite{brown2022voxsrc, huh2023voxsrc}, CHiME-6 \cite{watanabe2020chime} and Alimeeting \cite{yu2022m2met} employ such approach at least as part of a larger ensemble model \cite{wang2021dku, wang2022dku, wang2022similarity, wang2022cross}.
TS-VAD aims at producing speech probabilities for each speaker given a sequence of speaker-discriminative features e.g., in the original proposal \cite{medennikov2020target}, i-vectors of each speaker. 
As such, it can be considered more as a post-processing technique rather than a standalone diarization system since, to gather i-vectors or other speaker-id discriminative features, it is necessary first to perform  diarization via e.g., a clustering-based approach. After this initial diarization (which can be rather rough) the TS-VAD can be run iteratively to refine the speaker-id discriminative features and predictions simultaneously. 
Despite the obvious advantages, it can be argued that most TS-VAD methods have also some drawbacks, especially when actual industrial applications are considered. To be fair, some disadvantages are being addressed at the time of this writing and this is a very active research direction.  
For example, a recent work \cite{wang2022incorporating} expanded the TS-VAD approach to also incorporate the initial diarization step, obtaining thus a full-fledged standalone end-to-end diarization system (E2E TS-VAD), with impressive results on DIHARD-III.
Again, an issue is that most proposed TS-VAD approaches are offline, as the requirement for an initial diarization to gather speaker identities across the whole recording is best done offline. 
A notable exception is \cite{wang2022online} where a fully streamable TS-VAD method is proposed. Here the initial diarization is done in a chunk-wise fashion and the whole system becomes streamable, with competitive results with respect to a fully offline model \cite{wang2022cross}, especially for the two-speaker case.
Another possible drawback is the fact that many of the TS-VAD approaches that perform best are not flexible regarding the number of output speakers. In fact, for example in the original TS-VAD work \cite{medennikov2020target}, the best results are achieved for the system that models jointly all four speakers (by concatenating the i-vectors of each speaker) and the single, per-speaker system performs rather poorly compared to this one. 
Again, recent works are addressing this issue, for example, in \cite{cheng2022target}, a sequence-to-sequence TS-VAD approach is proposed, that can seamlessly generalize to arbitrarily output speakers and also different output resolutions.}

\textcolor{black}{UIS-RNN \cite{zhang2019fully} and DNC \cite{li2021discriminative} consist of supervised neural models based on speaker embeddings, where the first is clustering-free, whereas the other relies on neural-based clustering.
They are both not overlap-aware, but they are able to manage a variable number of speakers.}

Building on the invention of \ac{PIT} for DNN-based speech separation in 2019 \cite{kolbaek2017multitalker}, Fujita et al. developed the first fully end-to-end DNN-based system \cite{fujita2019end-blstm} which was later improved using self-attention \cite{fujita2019end-self}.
This again sparkled another research direction on systems based on end-to-end neural diarization (EEND) which is very active nowadays. 
Horiguchi et al. \cite{horiguchi2020end} proposed to extend EEND with an encoder-decoder-based attractor architecture (EEND-EDA) able to handle a flexible number of output speakers thanks to an autoregressive decoder.
\textcolor{black}{It embeds a speaker counting functionality consisting of an LSTM-based encoder-decoder applied on the EEND's output to generate a variable number of attractors from a speech embedding sequence (which is the EEND's output).
It also estimates the probability that the attractor is really associated with an active speaker.
It generates attractors until a stopping criterion is met (i.e., when the existence probability associated with an attractor is smaller than a threshold), thus it can deal with an arbitrary number of speakers.
Diarization results are obtained computing the dot product of the frame-wise speech embeddings with each speaker-wise attractor, followed by sigmoid activation. Speaker overlap is detected when multiple higher dot products between embeddings and attractors are present in the same frame \cite{horiguchi2022encoder}.}
Kinoshita et al. proposed several improvements to the initial EEND approach by integrating speaker-id embeddings extraction so that the strengths of EEND and clustering-based approaches can be combined \textcolor{black}{in a framework called EEND-vector clustering (EEND-VC)} \cite{kinoshita2021integrating, kinoshita2021advances, kinoshita2022utterance, kinoshita2022tight}.
\textcolor{black}{As EEND-EDA, EEND-VC is able to estimate the number of speakers, which can be arbitrary.
After splitting the input into chunks, it applies end-to-end diarization on each of them and derives speaker embeddings.
For each chunk the upper limit of the number of speakers is defined by the neural architecture, as for SA-EEND.
Chunk size has to be sufficiently long to obtain good speaker embeddings, whereas it has to be also sufficiently short to contain a number of speakers that does not exceed the maximum number allowed by the architecture.
Finally, the diarization output is obtained by applying a constrained-clustering algorithm between embeddings \cite{kinoshita2021integrating}.}
Other recent works focused on streaming processing: \cite{xue2021online} proposed to extend EEND with a speaker-tracing buffer to solve the permutation ambiguity caused by PIT when the model inference is performed via sliding windows. 
EEND-EDA streaming versions have been also recently proposed \cite{han2021bw, horiguchi2022online}. 
Also focusing on online processing, \cite{coria2021overlap}, combined the use of EEND with an x-vector extractor and online clustering, where the EEND model is used to gate the representation before the x-vector statistical pooling layer, to extract per-speaker embeddings even in overlap regions. 
Recent results in the most popular diarization challenges indicate that EEND-based systems are increasingly competitive \cite{wang2021ustc, horiguchi2021hitachi, landini2021but} and nowadays surpass clustering-based methods. 

Finally, other end-to-end approaches, that do not rely on PIT, such as DIVE \cite{zeghidour2021dive} \textcolor{black}{and the joint speaker diarization/ASR system based on a recurrent neural network transducer (RNN-T) proposed in \cite{shafey2019joint}}, have been also recently proposed.

\subsection{Motivation and Previous Review Works}\label{sec:motivation}
Many works attempted to make a comprehensive ``snapshot'' of the current advances and previous, up the time of their writing, algorithms proposed for speaker diarization.
In 2006, Tranter et al.~\cite{tranter2006overview} wrote one of the very first reviews of diarization systems developed from the 1990s until the early 2000s, in the broadcast news and CTS domains.
Following, in 2012, the work of Anguera et al.~\cite{anguera2012speaker} reviewed the new advancements, focusing mainly on meeting conversations, such as ones from AMI.
Park et al.~\cite{park2022review} made recently a very comprehensive review of speaker diarization including more up-to-time methods, which are largely based on DNNs. They proposed a taxonomy of speaker diarization which divided the algorithms into various groups.
These aforementioned works presented a comprehensive review of the speaker diarization field but did not analyze in depth the performance and computational aspects of different algorithms on a common dataset. 

Here, on the other hand, we aim at providing quantitative guidance on the performance of various recent diarization systems in a common scenario. To the best of our knowledge, this work is the first experimental review on speaker diarization. As said, we limit our scope to the CTS domain because current techniques seem particularly promising in this commercially relevant domain. 

We consider eight diarization systems which belong to clustering-based, end-to-end, and hybrid approaches and evaluate them on four multi-language datasets.
In addition, we study the computational cost of different systems (i.e., processing time, memory footprint). This aspect was neglected by previous review works, but it is of paramount importance as computational requirements highly impact the financial and environmental \cite{schwartz2020green} costs in production environments.

\section{Diarization systems under study}\label{sec:section_2}
Most diarization systems can be divided into two categories: clustering-based and end-to-end methods. There are, however, as explained, some notable exceptions (e.g., TS-VAD, SSGD, etc.), which are not fully end-to-end methods as involve usually a two-step procedure to obtain diarization (whilst clustering-based methods involve many different steps which are separately optimized) and make extensive use of deep learning based approaches.

In this paper, we perform a comparative study of algorithms that belong to clustering-based, end-to-end, and SSGD methods.

Regarding clustering-based approaches, we include a total of five different algorithms which come from the state-of-the-art such as VBx \cite{landini2020bayesian}, x-vector extractor with spectral clustering \cite{park2019auto} and their overlapped-aware variants \cite{bullock2020overlap, raj2021multi}. We also included a method based on \textcolor{black}{AHC and BIC} which is representative of early diarization approaches. 

Regarding end-to-end methods, we considered self-attentive EEND (SA-EEND) \textcolor{black}{\cite{fujita2019end-self}} and EEND-vector clustering (EEND-VC) \textcolor{black}{\cite{kinoshita2021integrating, kinoshita2021advances}}. Among the other methods that do not fall into EEND or clustering-based frameworks, we consider SSGD.
The choice of SSGD \textcolor{black}{\cite{fang2021deep, morrone2022low-latency}} over TS-VAD \textcolor{black}{\cite{medennikov2020target}} is considered as it is more straightforward to use (does not require initialization via clustering-based diarization) and has a greater commercial interest as the separated signals can be fed to an ASR back-end directly.
For example, it has been shown to be particularly effective in the CTS domain recently \cite{morrone2022low-latency} with near state-of-the-art diarization and ASR results close to ones obtained with oracle speech separation. 

We describe these different diarization paradigms in more detail thereafter. 

\begin{figure}[h!]
    \centering
    \subfloat[General architecture of a clustering-based diarization system.]{
        \includegraphics[width=0.85\columnwidth]{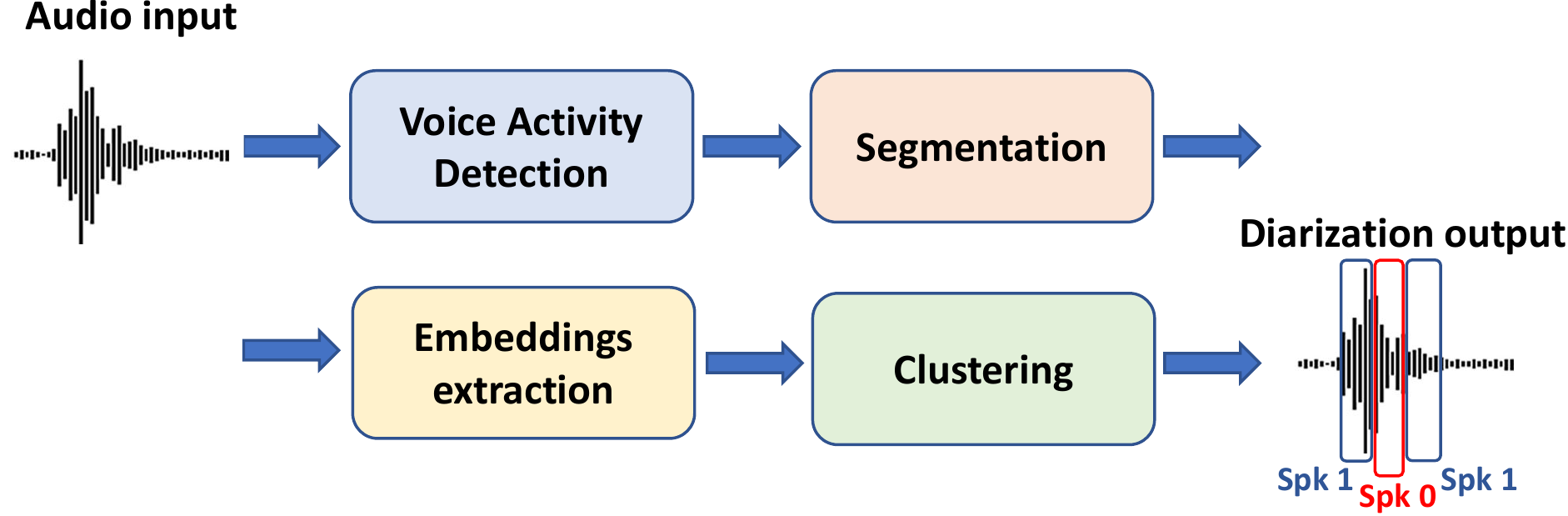}}
        \label{fig:clustering-based_architecture_V3}
    \vfill
    \vspace{0.2cm}
    \subfloat[Architecture of the SSGD system.]{
        \includegraphics[width=0.7\columnwidth]{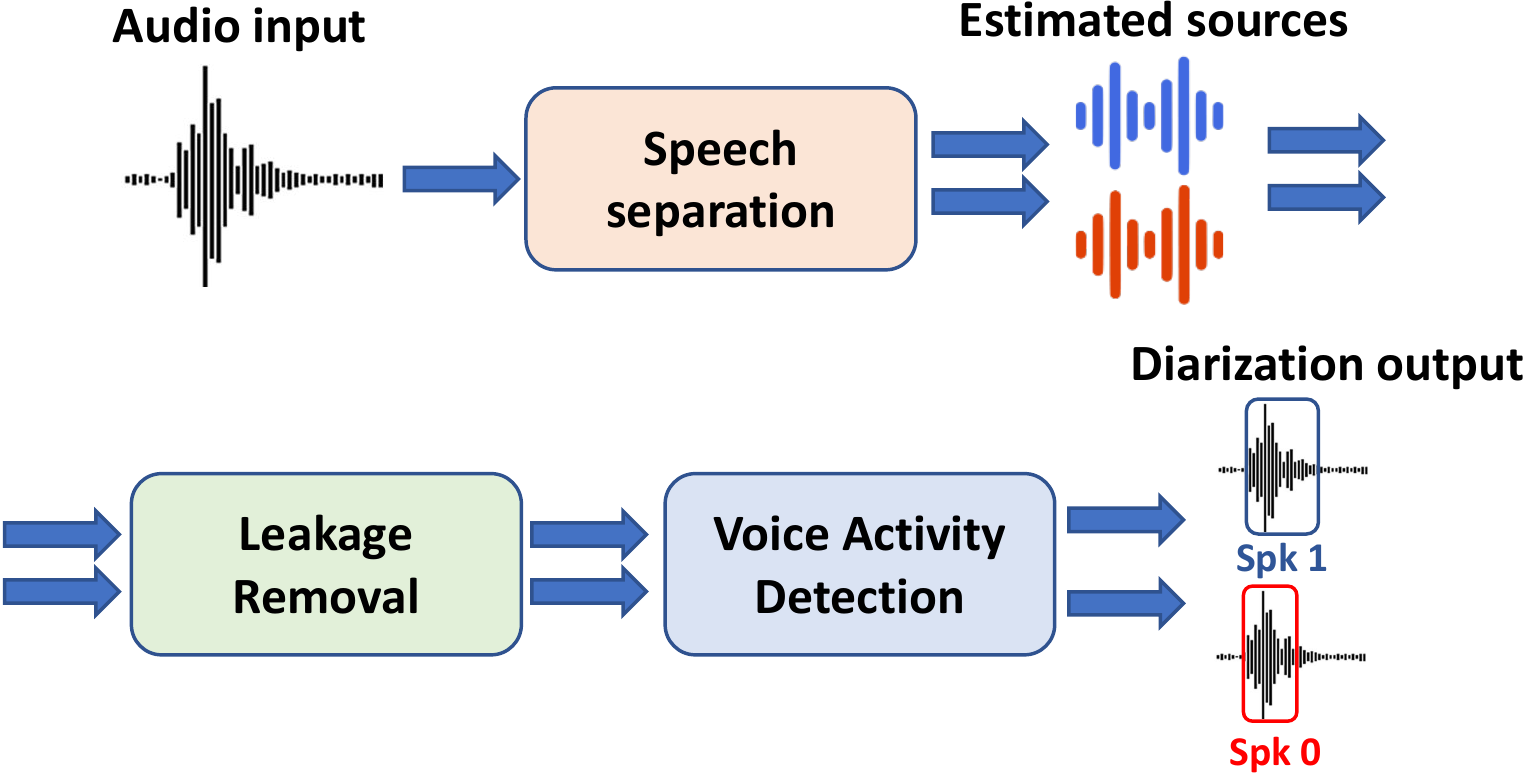}}
         \label{fig:SSGD_architecture_V1}
    \vfill
    \subfloat[Diarization system based on an end-to-end approach.]{
        \includegraphics[width=0.85\columnwidth]{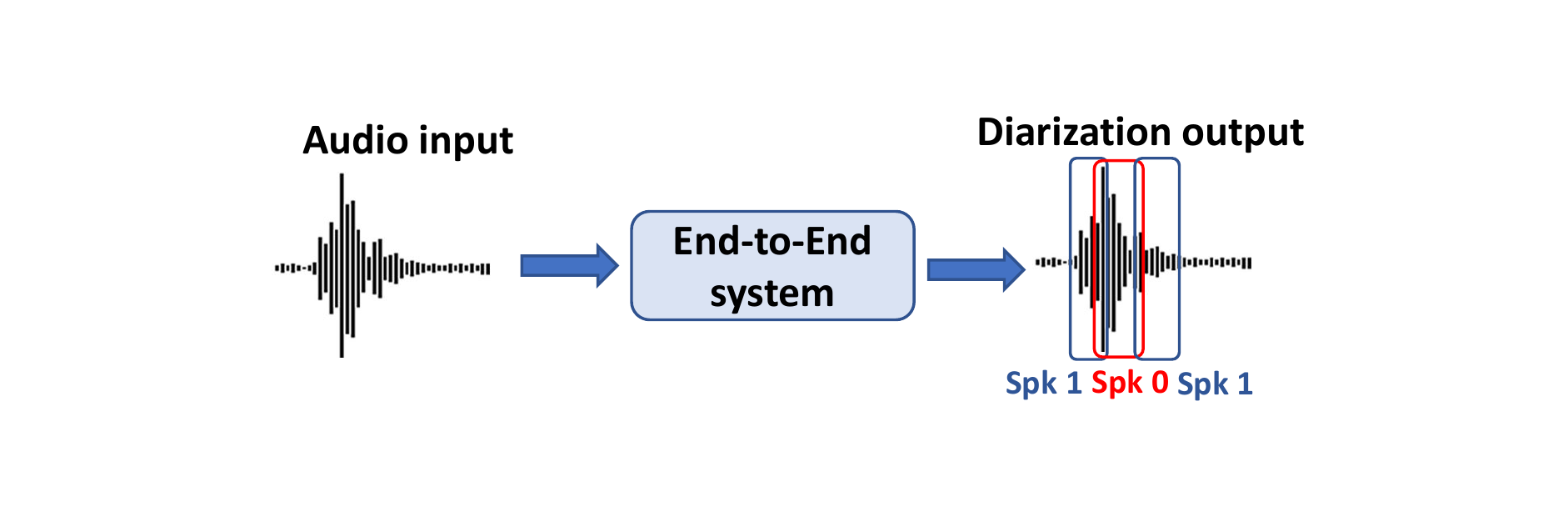}}
         \label{fig:EEND_general}
    \caption{General architectures of different categories of diarization systems.}
    \label{fig:overview_diarization_systems}
\end{figure}

\subsection{Clustering-based methods}\label{sec:clustering-based}

As explained in Section \ref{sec:history}, clustering-based methods have been the mainstream approach for diarization for several decades and continue to be relevant and competitive today. 
Clustering-based methods typically consist of four modules arranged in a pipeline structure (see Figure \ref{fig:overview_diarization_systems}(a)): VAD, segmentation, speaker embedding extraction, and clustering.

The VAD module detects portions of the audio signal containing voice only, by discarding non-speech parts, such as silence, music, background noise, etc. 
The objective of the segmentation module is to divide the speech portions into homogeneous single-speaker segments.
Early segmentation modules were based on the detection of speaker-change points, generating variable-length single-speaker segments.
In recent years, uniform segmentation gained success. It divides the input recording into overlapped fixed-length speech segments, which should be long enough to catch representative speaker information, without containing more than one speaker.
The extraction module's goal is to capture the speaker's characteristics within a speech segment and encapsulates them into a speaker embedding: a dense high-dimensional representation.
The clustering step groups the embeddings that belong to the same speaker.
Hereafter we describe the five clustering-based systems we considered in this experimental review.

\subsubsection{Variational Bayesian-Hidden Markov Model clustering of x-vector sequences (VBx)}
As mentioned in Section \ref{sec:history}, the VBx method proposed by Landini et. al.~\cite{landini2020bayesian} combines variational Bayesian resegmentation~\cite{diez2019analysis} with x-vector clustering diarization. It has been proven a highly effective diarization technique and this is attested by its wide use in popular diarization challenges such as DIHARD. 

VBx employs a standard clustering-based diarization pipeline based on x-vectors~\cite{snyder2018x} and PLDA+AHC clustering to derive initial diarization hypotheses. These are then refined via a Bayesian hidden Markov model (BHMM). 
In detail, the AHC step firstly derives hard assignments between speakers and the x-vectors sequence extracted from audio segments containing speech with a VAD. 
In VBx it is assumed that this sequence can be modeled by a hidden Markov model (HMM) with speaker-specific state distributions that are derived from the PLDA model. In detail, an ergodic HMM is used and each HMM state is associated with a single speaker, thus this approach cannot model overlapped speech. 
The hard assignment derived from AHC clustering is refined by finding the most probable sequence of HMM-associated latent variables that defines the hard alignment between the HMM states (and thus speakers) and x-vectors. This most probable sequence constitutes the diarization output of VBx and is found by firstly defining a hidden variable for the x-vector sequence distribution. This hidden variable is used to exploit the variational Bayes approach and the solution is found via evidence lower bound objective (ELBO) maximization.

\subsubsection{Spectral Clustering based on x-vectors (SCx)}\label{sec:spec_clust_x}
\textcolor{black}{The most used spectral clustering in early works on diarization was based on the NJW algorithm \cite{ng2001spectral}.}

In \textcolor{black}{NJW-based} spectral clustering, the speakers' clusters are estimated after the construction of a similarity matrix, a multi-dimensional vector whose elements represent pairwise similarities between speaker embeddings of different speakers.
This special matrix is transformed into spectral embeddings, which are then clustered through the k-means algorithm \cite{lloyd1982least}.

Its main drawback is represented by the limited accuracy since this technique is very sensitive to the noisiness of the similarity matrix due to intra-speaker variability of the speaker embeddings.  
In order to overcome this issue, a scaling parameter or a threshold should be introduced to assign different weights to pairwise similarities \cite{park2019auto}.
\textcolor{black}{This approach was followed, for example, by Wang et al. \cite{wang2018speaker}, which first exploited spectral clustering with maximum eigengap,} in place of the popular PLDA+AHC pipeline, to aggregate and cluster together DNN-extracted speaker embeddings, \textcolor{black}{i.e., d-vectors, and demonstrated it as state-of-the-art.}
The optimal values of these hyper-parameters have to be found through a tuning on the development set. This, however, can increase the computational cost and makes generalization more challenging.

\textcolor{black}{Shortly after, }Park et. al. \textcolor{black}{\cite{park2019auto}} propose to overcome this issue by deriving the optimal hyper-parameters through an auto-tuning strategy \textcolor{black}{that avoids the need for a development set.}

For simplicity, in this paper, \textcolor{black}{since we consider this last approach applied to x-vectors,} we will call it\textcolor{black}{, for convenience here,} SCx (in assonance to VBx).

\subsubsection{Overlap-aware versions of clustering-based diarization approaches based on x-vectors (VBx+OVL and SCx+OVL)}\label{sec:overlap_aware}
Conventional clustering-based approaches are unable to model overlap-speech and thus incur in severe diarization accuracy degradation in conversational speech. 
Recent works by Bullock et al. \cite{bullock2020overlap} and Raj et al. \cite{raj2021multi} proposed two different ways to make clustering approaches more ``overlap friendly" by leveraging an overlap detection front-end. 
In detail, \cite{bullock2020overlap} makes use of the VBx pipeline and combines it with an overlap detector. 
The output of VB resegmentation is a sequence of probabilities of each x-vector belonging to each speaker. 
In this overlap-aware algorithm, the most likely speaker sequence is multiplied by a mask obtained from the output of VAD module, whereas the second most likely sequence is masked with the overlap detection output.
In regions in which overlap has not been detected the output sequence is equal to the most likely speaker sequence. Instead, when overlap is detected, the output sequence is the union of the two most likely sequences.

The approach described in Raj et al. instead relies on the spectral clustering approach described previously in Section \ref{sec:spec_clust_x}.
Firstly, the authors propose to cast the discrete spectral clustering problem to continuous space via convex optimization. The continuous space then is discretized again, to obtain the speakers' assignments, but by using the output of an overlap detector to relax cluster exclusivity constraints.

\subsubsection{\protect\textcolor{black}{Clustering-based diarization algorithm using Agglomerative Hierarchical Criterion and the Bayesian Information Criterion (AHC-BIC)}}
The BIC can be used to perform clustering in a diarization system as outlined in \cite{chen1998speaker, tritschler1999improved}. 
The underlying assumption for BIC-based clustering is that the speaker-id discriminative features can be described by a multivariate Gaussian distribution. Early works such as \cite{chen1998speaker, tritschler1999improved} employed hand-crafted acoustic features such as MFCC which only approximately satisfy this assumption.
Once the features are collected these are merged iteratively \cite{tritschler1999improved} in an AHC way by computing the BIC \textcolor{black}{distance, $\Delta$BIC,} between each \textcolor{black}{pair of} cluster\textcolor{black}{s $i$ and $j$ as}:
\begin{equation}
    \label{delta_BIC}
    \Delta BIC = n \log |\Sigma| - n_i \log |\Sigma_i| - n_j \log |\Sigma_j| - \frac{\lambda}{2}\biggl\{d+\frac{d}{2}(d+1)\biggr\}\log n,
\end{equation}
where $|\Sigma|$ is the determinant of the covariance matrix estimated on acoustic data, $n = n_i + n_j$ (where $n_i$ and $n_j$ are, respectively, the number of segments belonging to cluster $i$ and $j$), $\lambda$ is a penalty factor that controls the aggressiveness of clustering (higher values lead to fewer clusters), and $d$ is the size of the acoustic vectors (the number of acoustic spectral features).

This $\Delta$BIC is a sort of penalized likelihood ratio expressing the difference between the assumptions that all the vectors are split into two Gaussian populations or belong all to the same Gaussian population, where the penalty is related to the different numbers of model parameters needed to represent the data in the two cases.
The smaller this value is, the closer the two clusters are, meaning that they are good candidates for merging.

\subsection{Speech Separation Guided Diarization (SSGD)}
In SSGD, diarization is performed through speech separation. 
As said, it was firstly proposed in \cite{siu1992unsupervised}, but has recently gained more attention \cite{fang2021deep, morrone2022low-latency}, as DNN-based methods for speech separation had become more and more effective. 
The classical SSGD pipeline is illustrated in Figure \ref{fig:overview_diarization_systems}(b).
A speech separation algorithm is firstly used to obtain from the input conversation audio estimated audio streams for each speaker. These audio streams are then fed to \ac{VAD} module which then estimates speech boundaries for each separated speaker, thus obtaining diarization annotation. 
As it happens with the VAD component in clustering-based systems, also here, usually post-processing, such as median filtering, is performed to smooth out the VAD predictions. 
In \cite{morrone2022low-latency} an additional ``leakage removal" post-processing module was proposed to reduce false alarms due to imperfect separation, occurring mainly on long single-speaker segments. 

It can be argued that, in SSGD, diarization almost comes ``for free", as the most computationally expensive operation is speech separation and the VAD component can be very simple. A simple energy-based VAD can already achieve decent performance \cite{morrone2022low-latency}. This makes it very appealing for use-cases that require downstream applications such as ASR or natural language understanding, for which we can use the available separated outputs.

\textcolor{black}{SSGD, as SA-EEND, needs a-priori knowledge of the maximum number of speakers in a conversation; therefore the system will work as desired even in single speaker situations, but not when the number of speakers is greater than the one considered during the training phase.
As SA-EEND, SSGD is able of operating in situations in which overlap is not present (overlap-free chunks).}

\subsection{End-to-End methods}
Clustering-based methods implicitly assume the presence of single-speaker segments only. Therefore, they are not able to handle situations where speaker overlap is present. As explained, overlap-aware assignment techniques such as \cite{bullock2020overlap, raj2021multi}, try to solve this problem but have intrinsic weaknesses, since the speaker embedding extractor module is still optimized without considering speaker overlap. Thus embeddings may not be very accurate on overlapped speech especially when overlap is significant. 

In contrast, end-to-end approaches can natively manage speech segments where multiple speakers are active at the same time. This makes them extremely appealing especially on domains such as CTS, where, as explained, the amount of overlapped speech can be quite significant. 

Their general structure is depicted in Figure \ref{fig:overview_diarization_systems}(c).
These methods have three main advantages over the clustering-based approaches.
First, they are conceptually simpler. Clustering-based algorithms need separate training for all the trainable models in the pipeline (i.e., VAD module, x-vector extraction module, PLDA, overlap detector). End-to-end systems require just one single model to be trained. 
Moreover, end-to-end systems can be trained to directly minimize diarization errors, in contrast to clustering which is usually performed in an unsupervised fashion.
Finally, whereas clustering-based systems rely on embedding extractors such as x-vectors that are trained for speaker verification on single-speaker segments only, end-to-end can be trained with real conversations, and thus naturally learn how to handle overlap and speaker turns. 

However, it can be argued that this last point could be also a weakness, as the cost of annotating real-world conversations, can constitute an obstacle to the adoption of end-to-end methods \cite{landini2022simulated}\textcolor{black}{; an evaluation study carried out by \cite{xia2022turn} quantifies this effort as approximately two hours of work to annotate 10 minutes of audio, for a human annotator.}
This cost can be reduced by using synthetically generated noisy mixtures as in \cite{fujita2019end-self, kinoshita2021advances}, however, care is needed during the training phase in order to avoid overfitting on a specific overlap ratio, and adaptation on a real dataset is then crucial to mitigate potential mismatches \cite{fujita2019end-self, horiguchi2020end, kinoshita2021advances}. 

In this work, we experiment with two end-to-End systems: SA-EEND and EEND-VC, both based on the PIT framework. We describe them thereafter.

\subsubsection{Self-Attentive End-to-End Neural Diarization \textcolor{black}{(}SA-EEND)}
SA-EEND \cite{fujita2019end-self} tackles the speaker diarization task by formulating it as a multi-label classification problem.
Diarization is performed within a single DNN, which is fed in with a sequence of acoustic features as input, and outputs the joint speech activities of multiple speakers i.e., VAD logits for each speaker.

SA-EEND is heavily inspired by DNN-based speech separation with PIT \cite{kolbaek2017multitalker}. 
The only difference from PIT-based speech separation is the fact that the output in EEND is the frame-wise speech presence posteriors for each speaker. 
As such, the standard binary cross entropy (BCE) loss is used, and the permutation ambiguity is tackled by selecting the speakers' permutation that minimizes this loss with respect to the current estimate. 
As it happens with speech separation it is necessary for SA-EEND that the number of output speakers is equal to or higher than the maximum number of speakers expected in one conversation. 
Also, the use of PIT means that the inference of the model should be performed in one pass over the conversation and not, sequentially on sliding windows as there is no guarantee that permutation will be consistent on different windows. 
A technique similar to continuous speech separation (CSS) \cite{chen2020continuous} can also be employed in practice, where outputs are re-ordered based on cosine-similarity with previous window predictions that overlap with current ones. 

\subsubsection{End-to-End Neural Diarization-Vector Clustering (EEND-VC)}
\begin{figure}[h]
\centering
\includegraphics[width=0.6\columnwidth]{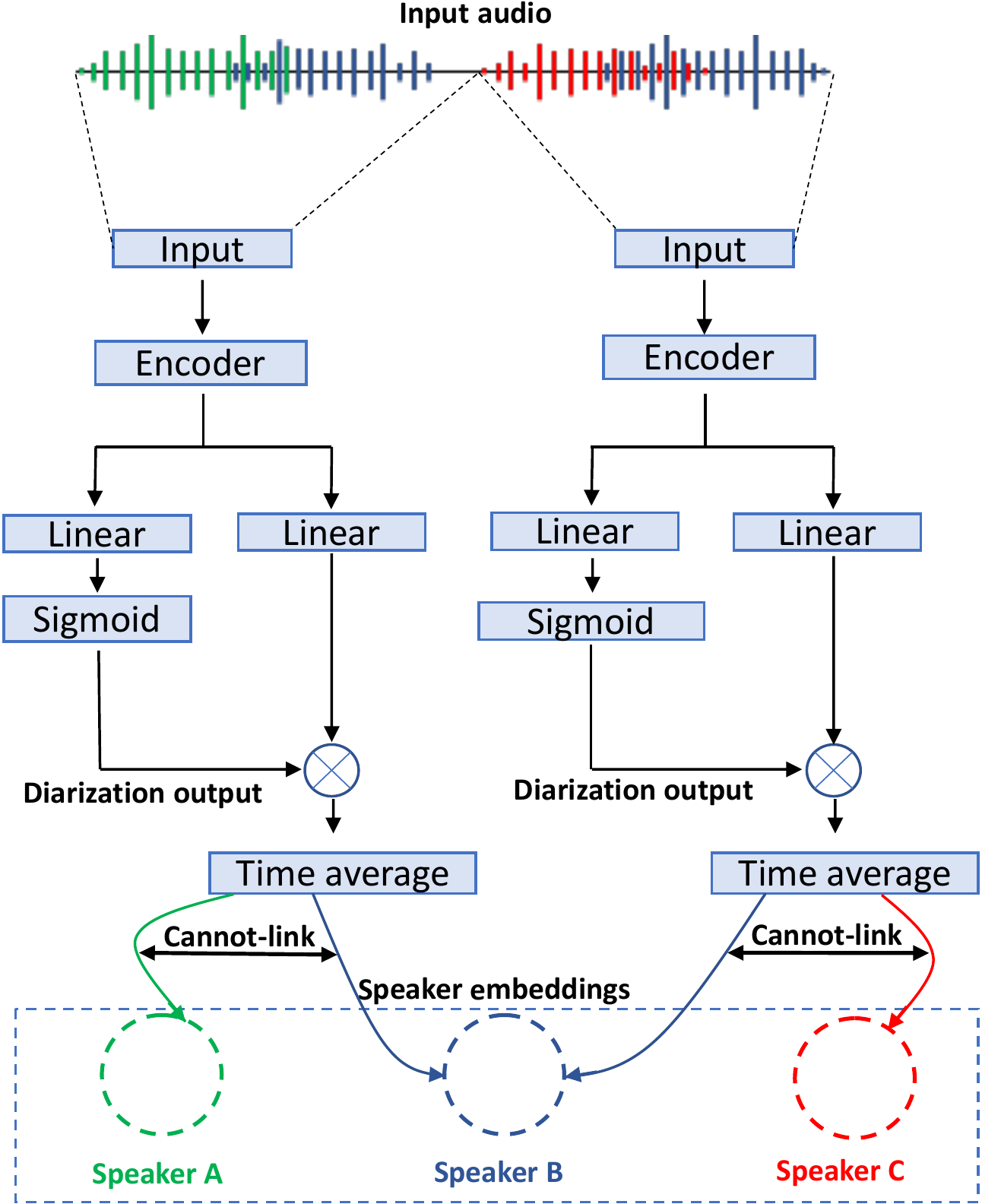}
\caption{Architecture of the EEND-VC.}
\label{fig:EEND-VC}
\end{figure}

This architecture \textcolor{black}{\cite{kinoshita2021integrating}} combines the EEND approach with clustering-based diarization principles in order to exploit the advantages of both methods.
It represents an improvement of the SA-EEND, as it is able to handle an arbitrary number of speakers and achieves better performance especially when input recordings are long.
As shown in Figure \ref{fig:EEND-VC}, it splits the audio input in chunks, and for each of them derives, as SA-EEND, local speakers' speech posteriors as well as their speaker embeddings associated with the current segment. 
In this way, the inter-chunk permutation problem of SA-EEND is avoided as the identity of speakers across a long conversation can be recovered via the clustering of speaker embeddings. This means that the model in inference can be run on chunks of several seconds, whose length is similar to the one employed in training, without the use of CSS-like stitching or a speaker buffer \cite{xue2021online}, which instead is necessary for SA-EEND. 
Another considerable advantage over SA-EEND is that EEND-VC only needs to know the maximum number of speakers within a single inference window which can be much smaller (e.g., 2, 3) than the total number of speakers in a conversation. 

EEND-VC can also leverage constrained clustering algorithms for preventing speaker embeddings belonging to the same chunk from being associated with the same speaker. This is in practice implemented by setting a cannot-link constraint between these embeddings. 
For example, original EEND-VC work~\cite{kinoshita2021integrating} used the COP-k-means \cite{wagstaff2001constrained} algorithm. In a follow-up work \cite{kinoshita2021advances} they demonstrated that the constrained AHC outperforms COP-k-means and other clustering algorithms, especially when the number of speakers is high.
When in one input chunk the number of speakers is less than the preset number of local speakers, it is necessary to identify whose speaker embeddings belong to silence. This is achieved in inference by looking at the local speakers' speech posteriors and seeing if their mean is below a certain ``activity threshold''. If this applies the corresponding embedding is considered as belonging to silence. 
During the learning phase, the network tries to minimize a loss function that consists of the weighted sum of two losses: the same PIT BCE loss as used in SA-EEND, and a speaker embedding loss.
The former is formulated in the same way as the SA-EEND loss function, but for local speakers, the latter is inspired by \cite{zeghidour2021wavesplit} and is designed to drive the speaker embeddings to have high inter-speaker distance and low intra-speaker variability.

\section{Experimental setup}\label{sec:section_3}

\subsection{Datasets}\label{sec:datasets}
In this experimental review, we consider four different two-speaker CTS datasets, all with 8\,kHz sample rate.  
Two of these are well-known, widely used datasets: CALLHOME~\cite{callhome}, Fisher~\cite{cieri2004fisher}. 
Another two datasets, i.e., CallCntrITA and CallCntrPOR, are novel and feature Italian and Portuguese languages, respectively.
The datasets have very different characteristics which are described in the following subsections. Statistics are shown in Table \ref{tbl_datasets}. We report the average total audio time, active speech time, speech sparsity \textcolor{black}{(computed as the amount of time where speech is not present divided by the total audio amount)}, overlap ratio, and audio quality using DNSMOS~\cite{reddy2021dnsmos} which is a blind mean opinion score (MOS) estimation method. 

\begin{table}[htpb!]
\scriptsize
\centering
\begin{tabular}{llllll}
\toprule
Metric & Unit & CALLHOME & Fisher & CallCntrITA & CallCntrPOR \\
\cmidrule{1-6}
Total audio time & h & 2.97 & 10.11 & 1.70 & 1.90 \\
Active speech time & h & 2.67 & 9.74 & 1.22 & 1.07 \\
Speech sparsity& \% & 10.\textcolor{black}{10} & 3.\textcolor{black}{6}6 & \textcolor{black}{28.24} & \textcolor{black}{43.68} \\
Audio file duration& s & 72.14 & 596.37 & 255.20 & 325.20 \\
Overlap ratio & $\%$ & 13.08 &14.59 &5.14 &5.35\\
DNSMOS  & - & 2.91  & 2.96 & 2.52 & 2.32 \\
\bottomrule
\end{tabular}
\caption{Main characteristics of the datasets. This table reports the average total audio time, active speech time, speech sparsity, overlap ratio (computed with respect to the active speech time duration), and DNSMOS of each dataset's test set.}\label{tbl_datasets}
\end{table}

\subsubsection{CALLHOME} 
CALLHOME \cite{callhome} is one of the most commonly used data corpora containing telephone conversations for speaker diarization evaluation.
It contains spontaneous conversations, distributed across six different languages: Arabic, English, German, Japanese, Mandarin, and Spanish.
Each conversation can have from two to a maximum of seven different speakers.

In this work, we use a two-speaker sub-portion of CALLHOME.
In particular, we use the subset of the 2000 NIST \ac{SRE} \cite{callhome} contained in the \verb+R65_8_1+ folder, as widely done in the literature, for example in EEND \cite{fujita2019end-blstm, fujita2019end-self}.
To allow for direct comparison, following such previous works, we select two-speaker conversations, and perform a split\textcolor{black}{, following the Kaldi CALLHOME diarization v2 recipe\footnote{\color{black}\url{https://github.com/kaldi-asr/kaldi/blob/master/egs/callhome_diarization/v2/run.sh}},} in order to obtain a validation set and a test set of 155 and 148 recordings, respectively.
As reported in Table \ref{tbl_datasets}, the CALLHOME test set is characterized by a relatively high overlap ratio, with respect to the other datasets in this study.

\subsubsection{Fisher Corpus Part I}\label{sec:fisher}
Fisher Corpus Part I \cite{cieri2004fisher} is an English-based dataset, containing 
telephone conversations between strangers on assigned topics. As such, dialogues are more formal than CALLHOME's ones.
Fisher provides separate streams for each of the two speakers; as separate streams are not available in many applications and also would make the diarization problem trivial, we sum these streams obtaining a mixture signal. In our experiments, we will perform diarization on these two-speaker mixture signals.
We split it in training (5\,000 files), validation (61 files), and test (61 files) partitions. 
As reported in Table \ref{tbl_datasets}, the test set of this corpora is the one with the largest overlap ratio and the smallest speech sparsity.
Therefore, its conversations are characterized by relatively few moments of silence and speakers often overlap. 

\subsubsection{CallCntrITA}
CallCntrITA is a proprietary data corpus that consists of real-world telephone conversations in a call-center scenario between a customer and the operator, in the Italian language. Audio files are recorded by a call center company that offers customer care oriented services.
The customer and operator are recorded in two different audio streams thus, as in Fisher, oracle speech sources are available. 
Again, as with Fisher, we sum them to obtain a mixture signal to be used as input to the diarization systems. 
CallCntrITA is composed of $48$ conversations which we randomly split in two equal validation and test partitions ($24$ each then).
As shown in Table \ref{tbl_datasets}, the associated test set presents the lowest overlap ratio among the datasets considered. Compared to CALLHOME and Fisher the DNSMOS values are lower because of the presence of babble noise on the operator \textcolor{black}{side}. CALLHOME and Fisher are instead relatively clean in most conversations as the high DNSMOS overall value suggests. 

\subsubsection{CallCntrPOR}\label{sec:data_por}
It is the Portuguese counterpart of CallCntrITA, thus it is recorded in the same manner.
It is composed of $42$ files and as for CallCntrITA, we randomly split it in two equal $21$ validation and test parts.
As shown in Table \ref{tbl_datasets}, the associated test set is denoted by a speech sparsity equal to $43$\%. thus, on average, almost half of the duration of each of its files consists of non-speech. It also presents one of the lowest overlap ratios.
It is the dataset with the lowest DNSMOS predicted audio quality: as CallCntrITA in the operator side there is babble noise but here it is more severe.

\subsection{Systems implementation details}\label{sec:impl_details}
Thereafter we report implementation details and hyper-parameters for the methods under study described in Section \ref{sec:section_2}: e.g., the SSGD's separation and speaker embedding extraction models employed in our experiments.
\textcolor{black}{Table \ref{tbl_systemscharacteristics} summarizes the major aspects regarding the implementation of all diarization systems considered in this manuscript (e.g., feature front-end, the ability to handle speakers overlap, training data) to highlight similarities and differences between them. More details are reported in each specific subsection, as follows.}

\subsubsection{Variational Bayesian-Hidden Markov Model clustering of x-vector sequences (VBx)}
For the VBx algorithm, we use an open source implementation\footnote{D. Raj, diarizer, \url{https://github.com/desh2608/diarizer/tree/is2022}\label{Desh_repo}} which was also used in \cite{morrone2022low-latency}. 
This implementation is derived directly from the official one made available by the VBx authors \cite{landini2020bayesian}.
For speech segmentation, we use the publicly available Kaldi ASpIRE \ac{VAD} model~\cite{peddinti2015jhu}\footnote{\url{https://kaldi-asr.org/models/m4}}\textcolor{black}{, which deals with 40 MFCC as input.
Since these are computed without dimensionality reduction, they contain the same information as filterbank features.}
As the x-vector extractor model we use a pre-trained 8\,kHz ResNet101 model\footnote{\color{black}\url{https://github.com/desh2608/diarizer/tree/is2022/diarizer/models/ResNet101_8kHz}} \cite{he2016deep}, and x-vectors are extracted every 0.24\,s, by considering a frame size equal to 1.44\,s.

\begin{figure}[htpb!]
\centering
\includegraphics[width=0.8\columnwidth]{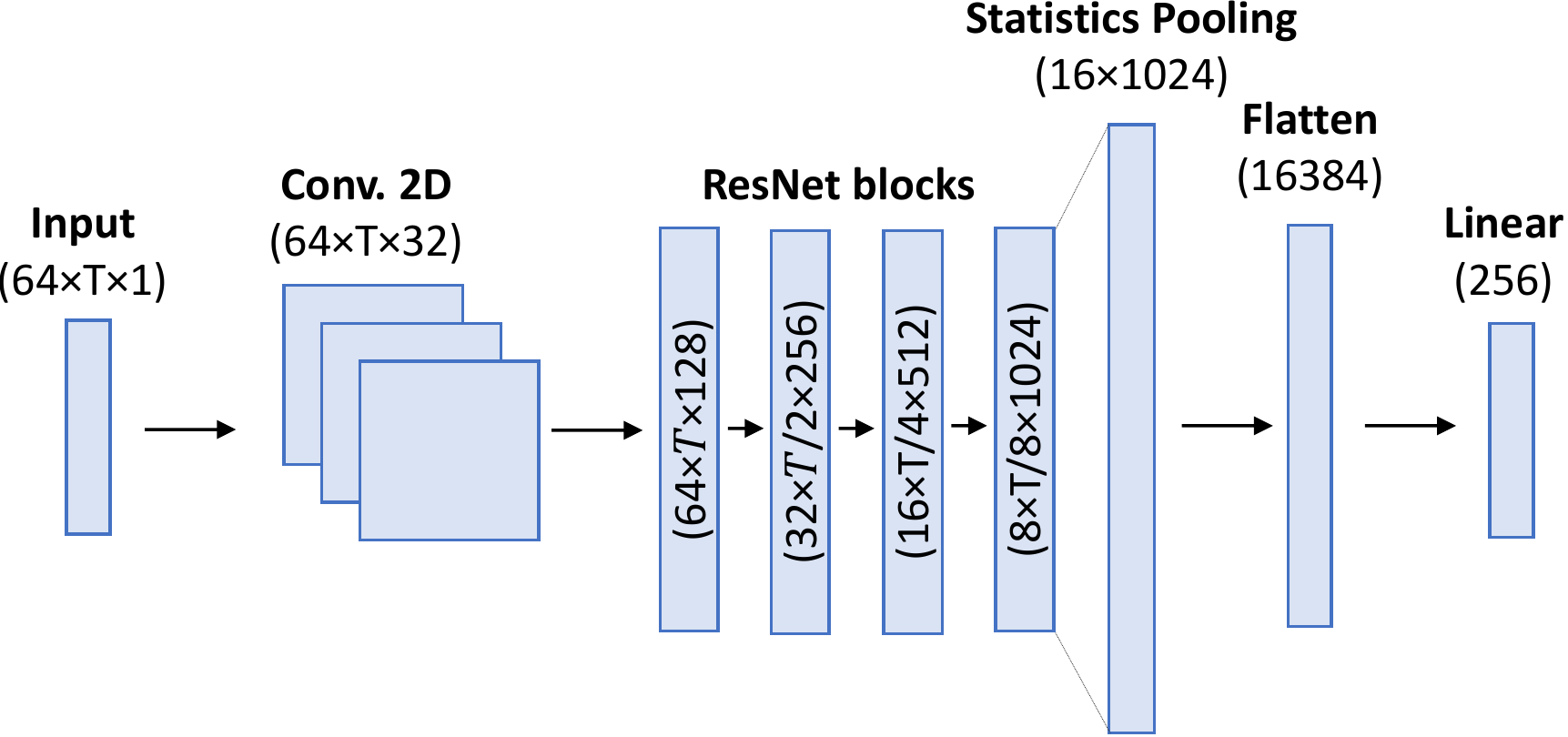}
\caption{Architecture of the x-vector extraction model used in the latest VBx system. The output size of each block is reported inside or near the respective box. $T$ indicates the number of input frames.}
\label{fig:x-vector_extractor}
\end{figure}

The ResNet101 x-vector extraction module is depicted in Figure \ref{fig:x-vector_extractor}. 
It consists of a neural network that takes $64$ \textcolor{black}{MFCC without dimensionality reduction} for $T$ frames as input.
Mel filters span the range 20-3\,700\,Hz, and features are extracted every 10\,ms by considering a window size equal to 25\,ms.
This model is trained for a speaker recognition task as the original x-vector architecture \cite{snyder2018x} on a large amount of data consisting of \textit{VoxCeleb1}
(323\,h, 1\,211 speakers) \cite{nagrani2017voxceleb}, \textit{VoxCeleb2} (2\,290\,h, 5\,994 speakers) \cite{chung2018voxceleb2}, \textit{CN-CELEB} (264\,h, 973 speakers) \cite{fan2020cn}, \textit{NIST-SRE 2004-2010} (3805\,h, 4\,254 speakers), \textit{Switchboard (SWBD)} (1\,170\,h, 2\,591 speakers), \textit{DeepMine} (688\,h, 1\,858 speakers) \cite{zeinali2018deepmine}. Data augmentation consisting of artificial reverberation and addition of music and noise from MUSAN~\cite{snyder2015musan} is also employed. \textcolor{black}{PLDA was trained on the same data}.

This same pipeline consisting of the Kaldi ASpIRE \ac{VAD} and the ResNet101 x-vector extractor, with their respective hyper-parameters (e.g., x-vector extractor hop and window) is also used in the other clustering-based techniques except for the \textcolor{black}{AHC-BIC}. 

The VBx hyper-parameters are tuned, for each dataset on the dataset development set. In detail, we tune VBx VB-HMM hyper-parameters: $F_A$, $F_B$, and $P_{loop}$.
The VBx implementation considered here uses PLDA+AHC for the initialization of the VB-HMM posterior matrix. The PLDA model is trained with the same data as the x-vector extraction module.

\subsubsection{Spectral clustering based on x-vectors (SCx)}
As previously reported in Section \ref{sec:section_2}, we also consider another clustering-based pipeline based on spectral clustering \textcolor{black}{with maximum eigengap, using an auto-tuning strategy} as proposed in \cite{park2019auto}\footref{Desh_repo}. Contrary to VBx, the resegmentation step here is not performed. We employ the same VAD model and ResNet-based x-vector extractor as the VBx one described in the previous section.

\subsubsection{Overlap-aware versions of clustering-based diarization approaches based on x-vectors (VBx+OVL and SCx+OVL)}
As explained in Section \ref{sec:overlap_aware}, the overlap-aware counterparts\footref{Desh_repo} of VBx and SCx are able to handle overlapped speech regions by leveraging an overlapped speech detector. Here we use a pre-trained
Pyannote segmentation model\footnote{\color{black} \url{https://github.com/desh2608/diarizer/blob/is2022/diarizer/models/pyannote/callhome_epoch4_step974.ckpt}} \cite{bredin2021end, bredin2020pyannote}.
\textcolor{black}{This model is trained on AMI dataset and, following \cite{morrone2022low-latency}, fine-tuned on the CALLHOME development set.}
\textcolor{black}{It uses 40 MFCC without dimensionality reduction, and} the hyper-parameters for this segmentation model are tuned on each dataset corresponding validation set. These include onset, offset, min\_duration\_on and min\_duration\_off.
Regarding the VAD module, the x-vector extractor, and the clustering module, we use the same implementations of VBx and SCx baseline versions.

\subsubsection{\protect\textcolor{black}{Clustering-based diarization algorithm using Agglomerative Hierarchical Criterion and the Bayesian Information Criterion (AHC-BIC)}}\label{sec:bic_implementation}
We used a proprietary diarization algorithm developed for call-center applications. It is highly optimized to run on resource-constrained devices.

As speaker-id discriminative features, it uses $12$ MFCC plus log-energy, extracted every 10\,ms by using a Hamming window 20\,ms long.
A proprietary energy-based \ac{VAD} based on \cite{junqua1994robust} is used to detect speech regions and reject silence. The VAD has been tuned to minimize false alarms by adjusting the threshold and putting a constraint on the minimum duration of segments labeled as speech. If this minimum duration is not met, the segment is labeled as non-speech. 

\subsubsection{Speech Separation Guided Diarization (SSGD)}
In this study, we use the SSGD implementation from \cite{morrone2022low-latency}. 
We use as the speech separator the dual-path recurrent neural network (DPRNN) \cite{luo2020dual} and, as the \ac{VAD} component, a temporal convolutional network (TCN) based one presented in \cite{cornell2022overlapped, morrone2022low-latency}.
\textcolor{black}{The former uses learned filterbanks \cite{luo2019conv} as input, whereas the latter exploits 40 logarithmic Mel-scaled filterbank energies features (LFBE);} both models are trained on the Fisher \textcolor{black}{Corpus Part I} dataset using the same procedure and hyper-parameters as described in \cite{morrone2022low-latency} which also employed synthetically mixed, fully overlapped speech examples from Fisher to pre-train the speech separation model. 
We consider a non-causal DPRNN and include in the pipeline also the leakage removal algorithm proposed in \cite{morrone2022low-latency}. This is also the configuration that in \cite{morrone2022low-latency} gives the best overall performance.
Following this latter work and SA-EEND, a median filter is also employed to smooth the TCN VAD predictions. 
The VAD threshold value and median filter length are tuned for each dataset on their respective development set. 

\subsubsection{Self-Attentive End-to-End Neural Diarization (SA-EEND)}\label{sec:eend_impl}
In this work, we used an open-source SA-EEND Pytorch \cite{paszke2019pytorch} implementation\footnote{Xflick, EEND PyTorch, \url{https://github.com/Xflick/EEND_PyTorch}}. 
We used the same features and model parameters as \cite{fujita2019end-self} except that we employed $4$ transformer encoder layers instead of $2$. 
The dataset used to train this model consists of a combination of SWBD-2 (Phase II, III), SWBD Cellular (Part 1, Part 2), and NIST Speaker Recognition Evaluation (NIST-SRE) datasets. The overall amount of data consists of $26\,172$ mixtures (2\,215\,h with an overlap ratio equal to $3.7\%$).
The training was performed as in \cite{fujita2020end} with Adam optimizer \cite{kingma2015adam} using the learning rate scheduler presented in \cite{vaswani2017attention}, where the decay phase is preceded by $25\,000$ warm-up steps.
The batch size was set to $64$ and the length of each example fed to the model had a length of 50\,s. 
The final model is achieved by averaging the model parameters obtained during the last $10$ epochs and by fine-tuning on the CALLHOME adaptation set. 
As in \cite{fujita2019end-self} median filtering is used to smooth the output predictions. Inference is performed not on chunks but by using all the conversation, as this model is trained with PIT. 
For each dataset, in this study, we tune the logits threshold and the median filter length using their development set. 

\subsubsection{End-to-End Neural Diarization-Vector Clustering (EEND-VC)}
We employ the implementation\footnote{nttcslab-sp, EEND-vector-clustering, \url{https://github.com/nttcslab-sp/EEND-vector-clustering}} from \cite{kinoshita2021advances}.
In order to carry out the experiments, we use a pre-trained model from \cite{kinoshita2021advances}. 
This model is comprised of $6$ transformer encoder blocks and uses $8$ attention heads in each multi-head self-attention layer. It uses the same input features (23 LFBE) as SA-EEND\textcolor{black}{, and uses Constrained AHC as a clustering algorithm.}
It was trained on simulated noisy three-speaker mixtures generated from: SWBD-2 (Phase I $\&$ II $\&$ III), SWBD Cellular (Part 1 $\&$ 2), and the NIST-SRE datasets; then, it was adapted on CALLHOME. These synthetic mixtures are created following the procedure considered in \cite{fujita2019end-blstm} with $\beta = 10$.
As the mixtures contain three speakers, the model is configured to have a maximum of three output local speakers.
\textcolor{black}{As it can be noted, EEND-VC pre-trained model is trained with mixtures generated by using the same datasets considered for the pre-trained SA-EEND, plus SWBD-2 Phase I; therefore their training data differ. In Section \ref{sec:sa_eend_vs_eend_vc} we compare them when they are trained on the same dataset.}

Training is performed with the same batch size and learning rate scheduler employed by SA-EEND, and Adam optimizer.
During training, the model is fed with chunks of length 15\,s, instead of 50\,s-chunks used in the case of SA-EEND. In inference, we use chunks of 30\,s.

As done for SA-EEND, also for EEND-VC we tune the output threshold and median filter length for each dataset.
The ``activity threshold'' to detect ``silent speakers'' \cite{kinoshita2021integrating} is set equal to 0.05 for all datasets and is not tuned. 

\begin{sidewaystable}[htpb!]
\begin{threeparttable}[htpb!]
\scriptsize
\centering\color{black} \captionsetup{labelfont={black},textfont=black}
\tabcolsep=0.10cm
\renewcommand{\arraystretch}{1.5}
\begin{tabular}{lcccccccc}
\toprule
& \multicolumn{5}{c}{\textcolor{black}{Modules}}\\
\cmidrule{2-6}
\textcolor{black}{} &  \textcolor{black}{Feature} & \textcolor{black}{} & \textcolor{black}{Spk Embed.} & \textcolor{black}{Clustering} & \textcolor{black}{} &  \textcolor{black}{Overlap} &  \textcolor{black}{Trainable} & \textcolor{black}{}\\
\textcolor{black}{System} &  \textcolor{black}{Front-End} & \textcolor{black}{VAD} & \textcolor{black}{extractor} & \textcolor{black}{Algorithm} & \textcolor{black}{Post-Processing} & \textcolor{black}{Handling} & \textcolor{black}{modules} & \textcolor{black}{Training data}\\
\cmidrule{1-9}
\textcolor{black}{VBx} & \textcolor{black}{MFCC} & \textcolor{black}{Kaldi ASpIRE\tnote{a}} & \textcolor{black}{ResNet101\tnote{b}} & \textcolor{black}{PLDA+} & \textcolor{black}{VB-HMM} &  \textcolor{black}{\xmark} & VAD$^{\ddag}$, PLDA$^{\dag}$, & \textcolor{black}{VoxCeleb 1\&2$^{\dag}$, CN-CELEB$^{\dag}$,}\\
\textcolor{black}{} &\textcolor{black}{} & \textcolor{black}{} &\textcolor{black}{} &\textcolor{black}{AHC init.} &\textcolor{black}{} &   \textcolor{black}{} & \textcolor{black}{x-vector extractor$^{\dag}$} & \textcolor{black}{NIST-SRE 2004-2010$^{\dag}$,}\\
\textcolor{black}{} &\textcolor{black}{} & \textcolor{black}{} &\textcolor{black}{} &\textcolor{black}{} & \textcolor{black}{} &  \textcolor{black}{} & \textcolor{black}{} &\textcolor{black}{SWBD$^{\dag}$, DeepMine$^{\dag}$}, Fisher$^{\ddag}$\\[10pt]

\textcolor{black}{SCx} & \textcolor{black}{MFCC} & \textcolor{black}{Kaldi ASpIRE\tnote{a}} & \textcolor{black}{ResNet101\tnote{b}} & \textcolor{black}{Spectral} & \textcolor{black}{-} &   \textcolor{black}{\xmark} &  \textcolor{black}{VAD$^{\ddag}$,} & \textcolor{black}{VoxCeleb 1\&2$^{\dag}$, CN-CELEB$^{\dag}$,}\\
\textcolor{black}{} &\textcolor{black}{} & \textcolor{black}{} &\textcolor{black}{} &\textcolor{black}{Clustering} &\textcolor{black}{} &   \textcolor{black}{} &  \textcolor{black}{x-vector extractor$^{\dag}$} & \textcolor{black}{NIST-SRE 2004-2010$^{\dag}$,}\\
\textcolor{black}{} &\textcolor{black}{} & \textcolor{black}{} &\textcolor{black}{} &\textcolor{black}{} &\textcolor{black}{} &   \textcolor{black}{} &  \textcolor{black}{} & \textcolor{black}{SWBD$^{\dag}$, DeepMine$^{\dag}$}, Fisher$^{\ddag}$\\[10pt]

\textcolor{black}{VBx+OVL} &\textcolor{black}{MFCC} & \textcolor{black}{Kaldi ASpIRE\tnote{a}} &\textcolor{black}{ResNet101\tnote{b}} &\textcolor{black}{PLDA+} &\textcolor{black}{VB-HMM} &  \textcolor{black}{\checkmark} &  \textcolor{black}{VAD$^{\ddag}$, PLDA$^{\dag}$,} & \textcolor{black}{VoxCeleb 1\&2$^{\dag}$, CN-CELEB$^{\dag}$,} \\
\textcolor{black}{} &\textcolor{black}{} & \textcolor{black}{} &\textcolor{black}{} &\textcolor{black}{AHC init.} &\textcolor{black}{} &  \textcolor{black}{} &  \textcolor{black}{x-vector extractor$^{\dag}$,} & \textcolor{black}{NIST-SRE 2004-2010$^{\dag}$, SWBD$^{\dag}$,} \\
\textcolor{black}{} &\textcolor{black}{} & \textcolor{black}{} &\textcolor{black}{} &\textcolor{black}{} &\textcolor{black}{} &  \textcolor{black}{} &   \textcolor{black}{OVL Detector\tnote{$\star$, c}} & \textcolor{black}{DeepMine$^{\dag}$}, Fisher$^{\ddag}$, AMI$^{\star}$ \\[10pt]

\textcolor{black}{SCx+OVL} &\textcolor{black}{MFCC} & \textcolor{black}{Kaldi ASpIRE\tnote{a}} &\textcolor{black}{ResNet101\tnote{b}} &\textcolor{black}{Spectral} &\textcolor{black}{-} &  \textcolor{black}{\checkmark} &  \textcolor{black}{VAD$^{\ddag}$,} & \textcolor{black}{VoxCeleb 1\&2$^{\dag}$, CN-CELEB$^{\dag}$,} \\
\textcolor{black}{} &\textcolor{black}{} & \textcolor{black}{} &\textcolor{black}{} &\textcolor{black}{Clustering} &\textcolor{black}{} &   \textcolor{black}{} &  \textcolor{black}{x-vector extractor$^{\dag}$,} & \textcolor{black}{NIST-SRE 2004-2010$^{\dag}$, SWBD$^{\dag}$,} \\
\textcolor{black}{} &\textcolor{black}{} & \textcolor{black}{} &\textcolor{black}{} &\textcolor{black}{} &\textcolor{black}{} &  \textcolor{black}{} & \textcolor{black}{OVL Detector\tnote{$\star$, c}} & \textcolor{black}{DeepMine$^{\dag}$}, Fisher$^{\ddag}$, AMI$^{\star}$ \\[10pt]

\textcolor{black}{AHC-BIC} &\textcolor{black}{MFCC+} & \textcolor{black}{Proprietary} &\textcolor{black}{-} &\textcolor{black}{AHC} &\textcolor{black}{-} &  \textcolor{black}{\xmark} &  \textcolor{black}{-} & \textcolor{black}{-} \\
\textcolor{black}{} &\textcolor{black}{log-energy} & \textcolor{black}{energy-based} &\textcolor{black}{} &\textcolor{black}{} &\textcolor{black}{} &  \textcolor{black}{} &  \textcolor{black}{} & \textcolor{black}{} \\[10pt]

\textcolor{black}{SSGD} &\textcolor{black}{Learned Filterbanks,} & \textcolor{black}{TCN-based} &\textcolor{black}{-} &\textcolor{black}{-} &\textcolor{black}{Leakage Removal} &  \textcolor{black}{\checkmark} &  \textcolor{black}{DPRNN Speech} & \textcolor{black}{Fisher Corpus Part I}\\
\textcolor{black}{} &\textcolor{black}{LFBE} & \textcolor{black}{} &\textcolor{black}{} &\textcolor{black}{} &\textcolor{black}{Algorithm} &  \textcolor{black}{} &  \textcolor{black}{Separator, VAD} & \textcolor{black}{}\\[10pt]

\textcolor{black}{SA-EEND} &\textcolor{black}{LFBE} & \textcolor{black}{Intrinsic} &\textcolor{black}{-} &\textcolor{black}{-} &\textcolor{black}{Median filter,} &  \textcolor{black}{\checkmark} &  \textcolor{black}{Whole model} &\textcolor{black}{SWBD-2 (II,III), NIST-SRE,} \\
\textcolor{black}{} &\textcolor{black}{} & \textcolor{black}{} &\textcolor{black}{} &\textcolor{black}{} &\textcolor{black}{output threshold} &  \textcolor{black}{} &  \textcolor{black}{} &\textcolor{black}{SWBD Cellular 1\&2} \\[10pt]

\textcolor{black}{EEND-VC} &\textcolor{black}{LFBE} & \textcolor{black}{Intrinsic} &\textcolor{black}{-} &\textcolor{black}{Constrained} &\textcolor{black}{Median filter,} &   \textcolor{black}{\checkmark} & \textcolor{black}{Whole model} &\textcolor{black}{SWBD-2 (I,II,III), NIST-SRE,} \\
\textcolor{black}{} &\textcolor{black}{} & \textcolor{black}{} &\textcolor{black}{} &\textcolor{black}{AHC} & \textcolor{black}{output threshold} &  \textcolor{black}{} &  \textcolor{black}{} &\textcolor{black}{SWBD Cellular 1\&2} \\

\bottomrule
\end{tabular}
\caption{\textcolor{black}{This table summarizes the major aspects regarding the implementation of all diarization systems considered in this work. Dagger, double dagger, and star ($\dag, \ddag, \star$) are used for clustering-based systems based on x-vectors to associate, respectively, the pair PLDA and x-vector extractor, the VAD, and the overlap (OVL) detector with datasets used to train them.}}\label{tbl_systemscharacteristics}

\begin{tablenotes}
       \color{black}\item [a] \url{https://kaldi-asr.org/models/m4}
       \color{black}\item [b] \url{https://github.com/desh2608/diarizer/tree/is2022/diarizer/models/ResNet101_8kHz}
       \color{black}\item [c] \url{https://github.com/desh2608/diarizer/blob/is2022/diarizer/models/pyannote/callhome_epoch4_step974.ckpt}

\end{tablenotes}
\end{threeparttable}
\end{sidewaystable}

\subsection{Evaluation Metrics}
This subsection presents the evaluation metrics we used to assess the performance of the diarization systems under study and their computational costs. 

\subsubsection{Diarization Accuracy}
We measure the speaker diarization systems' accuracy by using the \ac{DER} which is the most common metric used to evaluate the difference between the output of a diarization system and a reference oracle segmentation. It quantifies the fraction of the time not correctly attributed to a speaker or non-speech and is calculated as the ratio: 

\begin{equation}
    DER = \frac{MS + FA + SC}{TOT}
\end{equation}
where the denominator $TOT$ is the total speech time, obtained by summing the speech time of each speaker. The numerator is the sum of the following quantities: 

\begin{itemize}
\item \textit{Missed Speech (MS)}:
it is represented by the amount of scored time that a system wrongly indicates as non-speech.

\item \textit{False Alarm (FA)}:
as opposed to MS, it represents the portion of scored time that a system hypothesizes as speech, while it is labeled as non-speech in the reference.

\item \textit{Speaker Confusion (SC)}:
it is the portion of scored time assigned to the wrong speaker.
It does not take into account the portion of overlapped speech not detected.
Contrary to the other two errors, this is not related to the VAD stage, but it is strictly a diarization error due to segmentation and clustering steps.
\end{itemize}
In our experiments, we consider a ``fair'' evaluation setup, which takes into account all the speech (even segments containing overlap) and involves the introduction of a 0.25\,s ``no score collar'' placed at each segment's extremity.
This collar-based evaluation is a common practice and allows to mitigate annotation inconsistencies which are inevitable. Precise utterance boundaries are difficult to define unambiguously: e.g., oracle segmentation can be ``tighter" or ``looser" depending on the dataset and the instructions given to annotators. 
The DER computation is performed using the NIST md-eval scoring tool\footnote{N. Ryant, dscore, \url{https://github.com/nryant/dscore}} version 22.

\subsubsection{Computational Cost}
We assess the computational cost of each diarization system under study, by performing inference on the test set, for each dataset, and on a single-file basis. This mimics an actual deployment scenario where diarization is usually performed on each conversation separately so the output is readily available.

In detail, inference was performed on the same machine on an Intel i9-10920X CPU @ 3.50 GHz.
In this work, we focus on CPU inference as it is the most common deployment platform for a commercial application that has to run on the customer side (it may be a powerful personal computer or an edge device) for privacy reasons. In such instances, access to a hardware accelerator such as a GPU is not guaranteed. 
\textcolor{black}{
We performed inference by ensuring that no other users were using the machine and no other processes were running. Moreover, in order to carry out the experiments under a standard condition, simulating client-side deployment, we limited the number of cores to eight units.
All the diarization systems evaluated in this paper are implemented in Python, in particular by exploiting PyTorch, except AHC-BIC which is developed in C++. Anyway, it is important to mention that Pytorch also relies on calls to high-performance routines developed in C++, C or Fortran. For diarization systems with a pipeline structure, we calculated the inference time of each module net of the model’s loading time; then, we summed up the intermediate inference times together in order to compute the final result associated with the whole pipeline.}
For each algorithm and dataset, we calculated the average inference time, the Real-Time Factor (RTF), defined as the inference time divided by the audio file duration, and the maximum memory (RAM) occupation, during the diarization of each audio file.
To get a reliable estimate for these figures, inference is repeated ten times for each audio file and in the following Section \ref{sec:results_discussion} we will report both the average value and the standard deviation.

\section{Results and Discussion}\label{sec:results_discussion}

\subsection{Performance Evaluation}

Table \ref{tbl_DERs} reports diarization performance in terms of DER, obtained by performing inference with all the diarization algorithms under study on the test set of each dataset. 
To better understand which algorithm performs better across the four different scenarios we also computed the macro-average DER obtained by averaging the DER values across the four datasets. 
For this latter, we report also the standard deviation as it could outline which system is more consistent across all scenarios and thus may be preferable.

We can see that, overall, the best performing algorithm is EEND-VC, with VBx+OVL coming second.
\textcolor{black}{AHC-BIC} achieves lower accuracy than the other approaches. Its best performance is on CallCntrITA where it is rather close to SSGD.
Overlap-aware clustering is able to considerably boost performance for VBx and SCx. This is evident on the datasets with high speaker overlap, i.e., Fisher and CALLHOME: VBx+OVL is the algorithm that performs best on Fisher among all.
\textcolor{black}{As already reported in Table \ref{tbl_systemscharacteristics}, the diarization systems based on DNN-based speaker embeddings employ the same x-vector extractor which is also trained with the same data. Therefore the difference in their performance is not related to the speaker embedding extraction stage.}

SSGD has the least consistent performance among all algorithms under study and its performance degrades considerably on CallCntrITA and CallCntrPOR. 
This degradation may be due to the fact that, as it relies on speech separation, it is very sensitive to training/inference domain mismatch. Since it has been trained on Fisher, it thus performs well on CALLHOME as it has similar characteristics (e.g., speech sparsity, overlap ratio, high DNSMOS).
Instead, as explained in Section \ref{sec:datasets}, CallCntrITA and CallCntrPOR have slightly different acoustical conditions and speech statistics and thus the separation component is more prone to fail, degrading the final diarization accuracy.

SA-EEND performs slightly better than SSGD overall. The difference is very pronounced in CallCntrITA where SA-EEND seems better able to deal with sparse speech than SSGD. 
Compared to the other methods, it does not perform well however on long average length datasets such as Fisher, because of the mismatch between training sequence and inference sequence lengths.
In fact, it appears that SA-EEND methods are more sensitive to this mismatch than separation-based ones, which instead suffer from sparsity but can generalize better to long-lasting sequences in inference \cite{morrone2022low-latency}. EEND-VC is less prone to these problems, as it performs inference on fixed-size windows and it instead relies on clustering of locally extracted embeddings to solve the permutation ambiguity.
As with the other models, the DER for SA-EEND increases significantly on CallCntrPOR, both due to challenging acoustic conditions, high average length, and sparsity. 

VBx is overall the algorithm that has the most consistent performance (i.e., lowest standard deviation) across the scenarios owing to its simplicity and strong prior assumptions of its resegmentation process.

\begin{table}[htpb!]
\setlength{\tabcolsep}{3.4pt}
\scriptsize
\centering
\begin{tabular}{lccccccccc}
\toprule
& \multicolumn{8}{c}{DER}\\
\cmidrule{2-9}
Dataset & VBx  &SCx &VBx+OVL &SCx+OVL &\textcolor{black}{AHC-BIC}&SSGD & SA-EEND& EEND-VC\\
\cmidrule{1-9}
CALLHOME&11.74  &14.69 &10.13 &14.24  &22.92 &9.69 &10.98& \textbf{8.14}\\
Fisher&9.75  &9.15 & \textbf{6.91} &7.01  &26.13 &7.14 &11.04&7.60\\
CallCntrITA&10.53 & 8.87 & 10.47 & 8.79  & 12.86 &12.19 & 7.85 & \textbf{7.65} \\
CallCntrPOR& 14.26 &16.13 & \textbf{14.18} &16.12 &26.08 &25.15 &21.94& 14.25\\
\cmidrule{1-9}
& \multicolumn{8}{c}{macro-average DER}\\
\cmidrule{2-9}
&$11.57$ &$12.21$ &$10.42$ &$11.54$ &$22.00$ &$13.54$ &$12.95$ & $\mathbf{9.41}$\\
&$\mathbf{\pm1.71}$ &$\pm3.24$ &$\pm2.57$ &$\pm3.75$ &$\pm5.43$ &$\pm6.94$ &$\pm5.35$ &$\pm2.80$\\
\bottomrule
\end{tabular}
\caption{DER obtained by the different diarization algorithms on each test set.
For each diarization system, the table also reports the macro-average DER, which consists of the arithmetic mean (reported along with the associated standard deviation in the last row) of the DERs obtained on the different datasets. The best value in each row is highlighted in bold. 
}\label{tbl_DERs}
\end{table}

For a more complete analysis, in Table \ref{tab:miss_fa_conf} we report all the DER components: missed speech (MS), false alarms (FA), and speaker confusions (SC) for each dataset and each system under study. 
As expected, since they rely on the same VAD model, VBx and SCx have always the same MS and FA and differ only in terms of SC, as the clustering step is different and SCx lacks the VB-resegmentation step. 
We can also see that their overlap-aware modifications (i.e., VBx+OVL and SCx+OVL) improve the DER by reducing in particular MS and SC on CALLHOME and Fisher (the datasets with considerable overlap-ratio) while they slightly increase FA. 
The \textcolor{black}{AHC-BIC} pipeline always detects zero FA due to the fact that its energy-based VAD implements a strict constraint on the minimal utterance duration as said in Section \ref{sec:bic_implementation}. Most of the errors come from SC as it employs hand-crafted speaker-id discriminative features which are likely not robust enough to intra-speaker variability.

SSGD obtains the worst (compared to other datasets) FA and SC on both CallCntr datasets due to the speech sparsity and the presence of background babble noise both of which are more pronounced in CallCntrPOR. The noise degrades the speech separation algorithm output which then causes the VAD to give false alarms.

Regarding EEND methods, SA-EEND always generates more SC than EEND-VC. This latter can rely on clustering for speaker tracking whereas the former cannot. In fact, the SC difference is more noticeable on the datasets with high average conversations length.

\begin{table}[htpb!]
\setlength{\tabcolsep}{4.8pt}
\scriptsize
\centering
\begin{tabular}{lcccccccccccc}
\toprule
Method & \multicolumn{3}{c}{CALLHOME} & \multicolumn{3}{c}{Fisher} & \multicolumn{3}{c}{CallCntrITA} & \multicolumn{3}{c}{CallCntrPOR} \\
& MS & FA & SC & MS & FA & SC & MS & FA & SC & MS & FA & SC \\
\midrule
 VBx & 8.28 & \textbf{0.87} & 2.59& 8.47& \textbf{0.46} & 0.82& 2.88&5.19 &2.46 &4.41 &7.78 & \textbf{2.07} \\   
 SCx &8.28& \textbf{0.87} & 5.54 &8.47 & \textbf{0.46} &\textbf{0.22} &2.88 &5.19 & \textbf{0.80} &4.41 & 7.78& 3.93 \\
 VBx+OVL &5.33 & 2.36 & 2.44 & \textbf{3.95} &2.19 &0.77 &2.58 &5.44 &2.46 &4.33 &7.78 & \textbf{2.07} \\ 
 SCx+OVL &5.69 & 2.52 & 6.04 & \textbf{4.80} &2.02 & \textbf{0.19} &\textbf{2.05} &5.88 &0.86 &4.32 &7.78 &4.02 \\
 \textcolor{black}{AHC-BIC} & 6.39 & \textbf{0.00} &16.53 &8.20 & \textbf{0.00} &17.94 & 2.81 & \textbf{0.00} &10.05 & \textbf{2.16} & \textbf{0.00} &23.92 \\
 SSGD &5.92 &2.84 & \textbf{0.94} &4.86 &0.72 &1.56 & \textbf{0.41} &9.99 &1.79 & \textbf{2.34} &15.62 &7.20 \\
  SA-EEND & \textbf{4.66} &4.40 &1.92 &7.32 & 1.66 & 2.05 & 2.15 & 4.98 & 0.72 &5.94 & 12.57 & 3.43\\ 
 EEND-VC & \textbf{3.79} &3.32 & \textbf{1.03} &6.72 &0.62 & 0.26 &2.36 & \textbf{4.95} & \textbf{0.34} &6.99 & \textbf{6.29} & \textbf{0.97} \\

\bottomrule
\end{tabular}
\caption{DER numerator components: missed speech (MS), false alarms (FA) and speaker confusion (SC) for each diarization system and each dataset under study.
On each column we highlight in bold the best and second best values.}\label{tab:miss_fa_conf}
\end{table}

\begin{figure}[h!]
    \centering
    \subfloat[Pearson correlation coefficient (PCC) between DER and speech sparsity.]{
        \includegraphics[width=0.6\columnwidth]{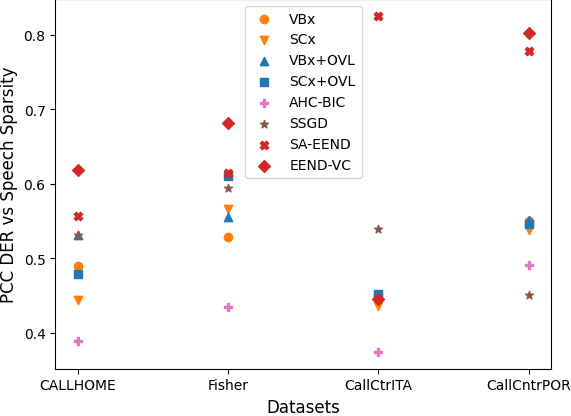}}
        \label{fig:PCC_DER_vs_sparsity}
    \vfill
    \subfloat[Pearson correlation coefficient (PCC) between DER and overlap ratio.]{
        \includegraphics[width=0.6\columnwidth]{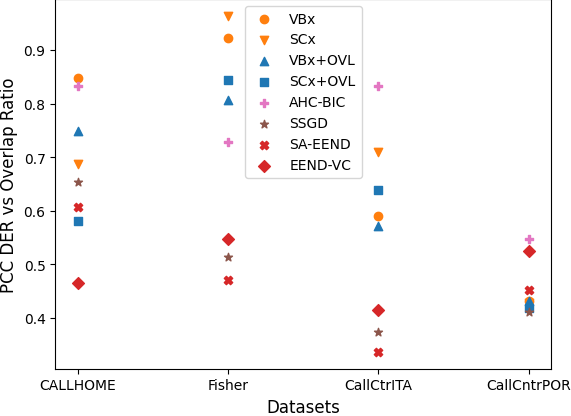}}
    \caption{Pearson correlation coefficient (PCC) between DER and speech sparsity (a) and overlap ratio (b) for each dataset and diarization method.}
         \label{fig:PCC_DER_vs_overlap}
    \label{fig:PCC_vs_sparsity_overlap}
\end{figure}

Figure \ref{fig:PCC_vs_sparsity_overlap} reports the Pearson correlation coefficient (PCC) between DER and speech sparsity (Figure \ref{fig:PCC_vs_sparsity_overlap}(a)) and between DER and overlap ratio (Figure \ref{fig:PCC_vs_sparsity_overlap}(b)) for each dataset and for each diarization method under study. The PCC was obtained by considering each recording in each dataset a different data-point: we computed the DER separately for each recording and used the speech statistics for each recording to compute the correlation.

Regarding correlation with speech sparsity, we can see that SA-EEND is the system that overall correlates the most. This is expected as it can struggle to correctly track speakers (due to PIT) when the conversation is sparse.
However, also EEND-VC exhibits some correlation, which is rather unexpected. This can be explained by the fact that EEND-VC is trained on a dataset with much less speakers' diversity than VBx and SCx (which use VoxCeleb 1 and 2). This means that it suffers more when the speech is very sparse as there are fewer segments to cluster and thus more intra-speaker variability due to less speaker diversity in training. Moreover, when the speech is sparse EEND-VC may not be able to exploit the cannot-link mechanism in the clustering step as only one speaker can appear in a particular signal window.
Also, SSGD seems to suffer from sparsity on all datasets but CallCntrPOR where it exhibits the poorest correlation among all systems indicating that the performance degradation on CallCntrPOR is due to the more challenging acoustical conditions and more pronounced babble noise compared to CallCntrITA. 

For what concerns the overlap ratio we can observe that, in general, the systems that suffer the most are the clustering-based ones without any overlap-aware mechanism (\textcolor{black}{AHC-BIC}, VBx, and SCx). 
As expected, their overlap-aware counterparts lower the correlation value. 
Overall, EEND and SSGD diarization systems are the ones that exhibit the lowest correlation across the datasets as they can process overlapped speech more effectively. 
On CallCntrPOR it seems that the performance, for most systems except \textcolor{black}{AHC-BIC} and EEND-VC, is largely unaffected by the overlap ratio.

\subsubsection{SA-EEND versus EEND-VC: Further Comparison}\label{sec:sa_eend_vs_eend_vc}
In addition to the analysis done before, we compare SA-EEND and EEND-VC further in a more fair setup: when both methods are trained from scratch on the same dataset \textcolor{black}{since, until this point, training data considered to obtain the EEND-VC pre-trained model differ from the ones used for SA-EEND pre-trained model e.g., for SA-EEND, SWBD-2 Phase I has not been taken into account.}
It makes sense to further compare the two EEND methods as their performance is highly dependent on the availability of large amounts of annotated conversations. Conversely, x-vector-based methods largely depend on the availability of datasets of pre-segmented, noisy utterances coming from a large number of different speakers to train the speaker-id feature extractor. Data for these latter do not need strict per-speaker segmentation annotation and are thus arguably easier to obtain in a more cost-effective way. 

We train these two models, with the same hyper-parameters as explained in Section \ref{sec:eend_impl} on Fisher Corpus Part I using the same split as described in Section \ref{sec:fisher}. We also change the maximum number of local speakers (and thus the output layer size) from three to two, because in this work we consider only two-speaker scenarios, whereas the pre-trained model from  \cite{kinoshita2021advances} was trained by assuming a maximum of three local speakers.

Both models are trained for 300 epochs, and we report results obtained by performing inference with the model checkpoint that achieved the lowest validation loss during the training phase.
As done for the two pre-trained models, we tune the threshold and, for post-processing smoothing, the median filtering length on the validation set of each of the four datasets separately. 

\begin{table}[h!]
\centering\color{black} \captionsetup{labelfont={black},textfont=black}
\begin{subtable}{0.45\linewidth}
\resizebox{\linewidth}{!}{
\begin{tabular}{lcc}
\toprule
& \textcolor{black}{SA-EEND} &\textcolor{black}{EEND-VC}\\
\cmidrule{2-3}
\textcolor{black}{Dataset} & \textcolor{black}{(H=4, L=4)} & \textcolor{black}{(H=8, L=6)}\\
\cmidrule{1-3}
\textcolor{black}{CALLHOME}&\textcolor{black}{12.15} & \textcolor{black}{\textbf{\underline{11.95}}}\\
\textcolor{black}{Fisher}& \textcolor{black}{10.80} & \textcolor{black}{\textbf{5.84}} \\
\textcolor{black}{CallCntrITA}& \textcolor{black}{11.73} & \textcolor{black}{\textbf{\underline{11.12}}}\\
\textcolor{black}{CallCntrPOR}&\textcolor{black}{\textbf{19.83}} & \textcolor{black}{20.97} \\
\bottomrule
\end{tabular}
}
\caption{\textcolor{black}{Diarization results obtained on different test sets, by training both EEND on Fisher Corpus Part I; the number of heads (H) and layers (L) is equal to the one considered in Table \ref{tbl_DERs}.}}\label{tab:1}
\end{subtable}
\hfill
\begin{subtable}{0.45\linewidth}
\resizebox{\linewidth}{!}{%
\begin{tabular}{lcc}
\toprule
& \textcolor{black}{SA-EEND} &\textcolor{black}{EEND-VC}\\
\cmidrule{2-3}
\textcolor{black}{Dataset} & \textcolor{black}{(H=8, L=6)} & \textcolor{black}{(H=4, L=4)}\\
\cmidrule{1-3}
\textcolor{black}{CALLHOME}& \textcolor{black}{14.96} & \textcolor{black}{\textbf{12.54}} \\
\textcolor{black}{Fisher}& \textcolor{black}{13.81} & \textcolor{black}{\textbf{\underline{5.60}}} \\
\textcolor{black}{CallCntrITA}& \textcolor{black}{\textbf{11.42}} & \textcolor{black}{12.47}\\
\textcolor{black}{CallCntrPOR}& \textcolor{black}{29.70} & \textcolor{black}{\textbf{\underline{19.79}}}\\
\bottomrule
\end{tabular}
}%
\caption{\textcolor{black}{Diarization results obtained on different test sets, by training both EEND on Fisher Corpus Part I; the number of heads (H) and layers (L) for SA-EEND is the ones used for EEND-VC in Table \ref{tbl_DERs}, and vice versa for EEND-VC.}}\label{tab:2}
\end{subtable}
\caption{\textcolor{black}{Diarization results obtained on different test sets, by training both EEND on Fisher Corpus Part I.
In (a) the number of heads (H) and layers (L)  is equal to the default ones considered in Table \ref{tbl_DERs}, whereas results in (b) are computed by swapping the (H, L) pairs of the two EEND. For each subtable, the best result on each test set is reported in \textbf{bold}, whereas the best values across subtables are \underline{underlined}.}}\label{tbl_comparison}
\end{table}

All results are reported in Table \ref{tbl_comparison}a. For EEND-VC we suppose that the total number of speakers is known a-priori and we exploit this information during the clustering phase. 
As expected, EEND-VC outperforms SA-EEND on most datasets, whereas SA-EEND slightly outperforms EEND-VC on CallCntrPOR. 
EEND-VC is significantly better on Fisher, again, as its reliance on clustering for speaker tracking allows for better generalization to longer sequences even if it has been trained on shorter ones. 

Comparing Table \textcolor{black}{\ref{tbl_comparison}a} with Table \ref{tbl_DERs} \textcolor{black}{, which both consider (H=4, L=4) for SA-EEND and (H=8, L=6) for EEND-VC}, we can notice that the performance of EEND-VC decreases for all datasets except Fisher, where instead it improves due to the perfectly matched training and test. 
For SA-EEND we observe the same except that now the performance is also slightly better for CallCntrPOR.

\textcolor{black}{In general, the performance loss by training on Fisher Corpus Part I, which is a much smaller dataset than the ones used for both EEND systems in Table \ref{tbl_DERs}, is quite significant.
Overall this result suggests that both these approaches are especially data hungry, as the dataset used here is not that small, and that state-of-the-art results are possible only when a large amount of data are available (e.g., several thousands of hours, as used in EEND-VC and SA-EEND original works).}

Finally, we can analyze how these two EEND-based approaches trained on Fisher compare with the SSGD-based pipeline which was also trained using Fisher and whose results were reported in Table \ref{tbl_DERs}.
SSGD is now more competitive with respect to EEND-VC but overall this latter still affords lower DER on most datasets, especially on CallCntrPOR and Fisher. 
Both EEND-based approaches also appear more robust than SSGD on the two CallCntr datasets which have mismatched acoustic characteristics and different speech statistics compared to Fisher and CALLHOME. SSGD is more sensitive to such a mismatch. 

\textcolor{black}{In order to also carry out a comparison in which their number of parameters is the same, we trained both EEND models on Fisher Corpus Part I, setting for SA-EEND the same number of heads (H) and layers (L) used to obtain the results of EEND-VC in Table \ref{tbl_comparison}a, and vice versa for EEND-VC.
Then, SA-EEND was trained with H=8 and L=6, whereas EEND-VC with H=4 and L=4; we report these results in Table \ref{tbl_comparison}b.}

\textcolor{black}{Therefore, by comparing Tables \ref{tbl_comparison}a and \ref{tbl_comparison}b, we can analyze two situations in which the number of parameters of the two EEND is on par; in particular, instances with (H=4, L=4) and (H=8, L=6).}

\textcolor{black}{In the latter case, we can see that EEND-VC performs better than SA-EEND on all test sets.
For the former case, SA-EEND outperforms EEND-VC only on CALLHOME and CallCntrITA test sets, whereas is worse on Fisher and CallCntrPOR.}

\textcolor{black}{SA-EEND is the one that suffers the most from the usage of an (H, L) pair  different from the one considered in their pre-trained models.}

\textcolor{black}{Overall, it can be seen that the best EEND on all test sets still remains the pre-trained EEND-VC model of Table \ref{tbl_DERs}, thanks to the larger amount of data of the training set.
The same states even when these EEND models are trained on the same data (i.e., Fisher Corpus Part I) when their number of layers is not the same as they have, respectively, in Table \ref{tbl_DERs}.
Concerning experiments in which both models are trained on the same data and the number of layers on par, EEND-VC outperforms SA-EEND on all test sets when (H=8, L=6), and even on Fisher and CallCntrPOR when (H=4, L=4).}

\subsection{Computational Cost}
\begin{figure}[h!]
    \centering
    \includegraphics[width=0.7\columnwidth]{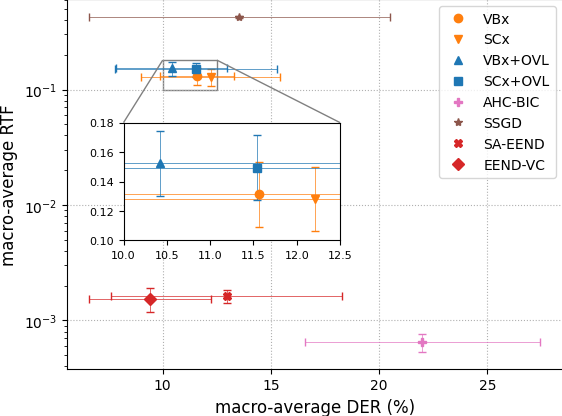}
    \caption{RTF versus DER for the algorithms in exam. For each, we report average values and standard deviations (using bars) across the four datasets in exam.}
    \label{fig:RTF_vs_DER}
\end{figure}

\begin{figure}[h!]
    \centering
    \includegraphics[width=0.7\columnwidth]{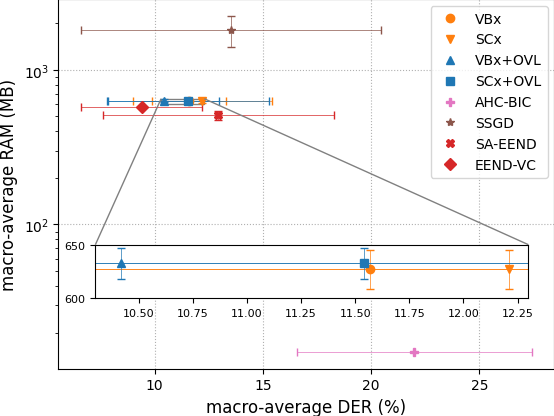}
    \caption{Maximum RAM occupation in Megabytes (MB) versus DER for the algorithms in exam. For each, we report average values and standard deviations (using bars) across the four datasets in exam.}
    \label{fig:RAM_vs_DER}
\end{figure}

\begin{figure}[h!]
    \centering
    \includegraphics[width=0.7\columnwidth]{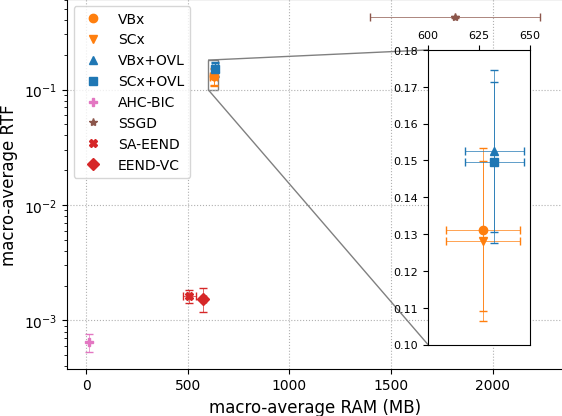}
    \caption{RTF versus RAM for the algorithms in exam. For each, we report average values and standard deviations (using bars) across the four datasets in exam.}
    \label{fig:RTF_vs_RAM}
\end{figure}

\textcolor{black}{In order to visualize the accuracy of each algorithm together with its performance from a computational perspective, we use Figures \ref{fig:RTF_vs_DER}, \ref{fig:RAM_vs_DER}, and \ref{fig:RTF_vs_RAM}; in order to have a more detailed overview, the precise values are also reported in Table \ref{tbl_DERs_RTFs_RAM}.}

In Figure \ref{fig:RTF_vs_DER}, we plot the RTF versus DER for each system under study. Both DER and RTF values have been averaged across the four datasets. The DER and RTF standard deviations are reported using bars.
We can observe that both EEND methods in general offer a very good trade-off between performance and processing time, with EEND-VC being the best overall. 
Clustering-based techniques have very similar RTFs overall and overlap-aware variants (VBx+OVL and SCx+OVL) only add a small computational overhead as the most time-consuming operation consists in the embedding extraction. 

SSGD has the highest RTF due to the fact that neural speech separation is quite expensive, especially as most current state-of-the-art models are largely based on learnt over-complete filterbanks \cite{luo2020dual, luo2019conv} with small kernels. On the other hand, as said, SSGD can be more convenient in all applications when it is necessary to transcribe both speakers, as diarization comes with a negligible computational overhead after separation. 
As expected, the \textcolor{black}{AHC-BIC} system is the fastest but its performance in terms of DER is also the worst amongst all. 

\textcolor{black}{From Figure \ref{fig:RTF_vs_DER} (and even better from Table \ref{tbl_DERs_RTFs_RAM}) it is possible to see that the standard deviation of the RTF (computed by averaging it across the four datasets) of DNN-based clustering systems is about 0.02, for SSGD is less than half, whereas for EEND systems and AHC-BIC is two orders of magnitude lower.
The low standard deviation of RTF means that inference time is approximately proportional to the audio file duration.
This aspect, in addition to indicating the low dependency of inference time on the different characteristics of each audio file (e.g., duration, speech sparsity, etc.), also reflects the reliability of the setup (no background tasks, etc.) used to conduct the experiments.
Thus, the slightly greater standard deviation of clustering-based systems could be due to their greater susceptibility to speech sparsity variability, since they rely on x-vectors, which are computed only on speech segments, after the VAD.}

In Figure \ref{fig:RAM_vs_DER}, we plot the maximum RAM occupation in inference versus DER for each system under study. Again for both we plot the macro-average across all datasets. 
We can see that most systems have very similar maximum RAM requirements: VBx, SCx, their OVL variants, and the two EEND models are under 1\,GB of RAM.
As expected, the SSGD model instead is the most memory-demanding due to the aforementioned fact about the filterbanks, which increases the amount of information the model needs to process.

Since it could be argued that, depending on the actual implementation, it is usually possible to trade-off speed (and thus here RTF) for memory occupation, in Figure \ref{fig:RTF_vs_RAM} we plot the RTF versus the maximum RAM occupation, again by macro averaging across the datasets.
It is evident that, for the time versus space complexity trade-off, the systems that lie closer to the origin offer intrinsically better computational parsimony. 
We can see that the \textcolor{black}{AHC-BIC} and the two EEND approaches are by far the most efficient, whereas SSGD is the least efficient. 
The clustering-based approaches are largely equivalent despite their differences, as, again, the bulk of the processing is due to the extraction of the x-vectors. 

\textcolor{black}{From Figure \ref{fig:RTF_vs_RAM}, it could be seen that SSGD is the system that suffers the most the variability between the different datasets, because its RAM's standard deviation is about 400 MB; in particular, the longer audio
duration the greater the RAM occupation. In contrast, DNN-based clustering systems and EEND models experience a RAM’s standard deviation of only a few tens of MB, and even lower for the AHC-BIC.
EEND-VC shows stronger robustness to file duration with respect to SA-EEND, as its RAM occupation is about three times lower.}

In general, we can say that EEND-VC affords the most favourable trade-off between performance and computational efficiency. SSGD achieves decent performance but is the least efficient.
The clustering-based methods, excluding \textcolor{black}{AHC-BIC}, offer similar computational trade-offs and thus, VBx+OVL, which performs slightly better overall, should be preferred.
\begin{table}[htpb!]
\setlength{\tabcolsep}{3.4pt}
\scriptsize
\centering\color{black} \captionsetup{labelfont={black},textfont=black}
\begin{tabular}{llccccccccc}
\toprule
& \multicolumn{9}{c}{\textcolor{black}{Diarization systems}}\\
\cmidrule{3-10}
\textcolor{black}{Metric} & \textcolor{black}{Unit} & \textcolor{black}{VBx} &\textcolor{black}{SCx} &\textcolor{black}{VBx+OVL} &\textcolor{black}{SCx+OVL} &\textcolor{black}{AHC-BIC} &\textcolor{black}{SSGD} & \textcolor{black}{SA-EEND} & \textcolor{black}{EEND-VC}\\
\cmidrule{1-10}
\textcolor{black}{DER} & & \textcolor{black}{$11.57$} &\textcolor{black}{$12.21$} &\textcolor{black}{$10.42$} &\textcolor{black}{$11.54$}&\textcolor{black}{$22.00$} &\textcolor{black}{$13.54$} &\textcolor{black}{$12.95$} & \textcolor{black}{$9.41$}\\
& & \textcolor{black}{$\pm$1.71} & \textcolor{black}{$\pm$3.24} & \textcolor{black}{$\pm$2.57} & \textcolor{black}{$\pm$3.75} & \textcolor{black}{$\pm$5.43} & \textcolor{black}{$\pm$6.94} & \textcolor{black}{$\pm$5.35} & \textcolor{black}{$\pm$2.80}\\
\textcolor{black}{RTF} & & \textcolor{black}{1.31e-1}  & \textcolor{black}{1.28e-1} & \textcolor{black}{1.52e-1} & \textcolor{black}{1.49e-1} & \textcolor{black}{6.45e-4} & \textcolor{black}{4.25e-1} &\textcolor{black}{1.64e-3} & \textcolor{black}{1.54e-3}\\
 & & \textcolor{black}{$\pm$2.20e-2}  &\textcolor{black}{$\pm$2.18e-2}&\textcolor{black}{$\pm$2.20e-2} &\textcolor{black}{$\pm$2.18e-2} &\textcolor{black}{$\pm$1.11e-4} &\textcolor{black}{$\pm$9.36e-3} &\textcolor{black}{$\pm$2.06e-4} & \textcolor{black}{$\pm$3.67e-4}\\
\textcolor{black}{RAM} & \textcolor{black}{MB} & \textcolor{black}{627} & \textcolor{black}{627} & \textcolor{black}{633} & \textcolor{black}{633} & \textcolor{black}{15} & \textcolor{black}{1814} & \textcolor{black}{508}& \textcolor{black}{577} \\
 & & \textcolor{black}{$\pm$18} & \textcolor{black}{$\pm$18} & \textcolor{black}{$\pm$15} & \textcolor{black}{$\pm$15} & \textcolor{black}{$\pm$0.02} & \textcolor{black}{$\pm$419} & \textcolor{black}{$\pm$31} & \textcolor{black}{$\pm$11}\\

\bottomrule
\end{tabular}
\caption{\textcolor{black}{DER, RTF, and RAM obtained by performing inference with the different diarization algorithms under study.
For each of them, we report the mean value, averaged on the four test sets, along with the associated standard deviation.
}}\label{tbl_DERs_RTFs_RAM}
\end{table}

\section{Conclusion}\label{sec:conclusion}
In this paper, we present an experimental review of speaker diarization in the CTS domain and investigate clustering-based, end-to-end, and separation-guided diarization methods.
In total, we considered eight different methods and analyzed their performance in terms of diarization accuracy and computational requirements using four different datasets.

Our goal is to give quantitative guidance concerning which diarization approach might be the most appropriate depending on the specific use-case scenario, e.g., the characteristics of the audio recordings and the computational capabilities of the machine that can be used for inference. 

Among the diarization methods, we considered VBx and a clustering-diarization pipeline based on spectral clustering (SCx), as well as their overlap-aware modifications (VBx+OVL, SCx+OVL)\textcolor{black}{,} plus \textcolor{black}{an approach based on AHC, BIC, and} hand-crafted features, denoted as \textcolor{black}{AHC-BIC}.
Regarding end-to-end methods, we tested SA-EEND and EEND-VC. The SSGD pipeline instead was taken from a recent work focused on low-latency diarization in the CTS domain. 
The datasets used are from multiple different languages and have different characteristics in terms of speech sparsity and presence of overlapped speech. We included both publicly available, widely used CTS datasets such as CALLHOME as well as two proprietary datasets from industry. 

Regarding the clustering-based approaches, we found that the computational burden of overlap-aware VBx and SCx are comparable with their non overlap-aware versions, as most resources are due to the x-vector extraction step. 
In low overlapped speech scenarios (e.g., call-center customer care), their performance is comparable but, as expected, overlap-aware versions bring noticeable improvement on the datasets with more overlapped speech featuring more colloquial speech. 
VBx and its overlap-aware variant (VBx+OVL) were found to perform the best overall among clustering-based techniques. 
The \textcolor{black}{AHC-BIC} approach has the worst performance overall among all the methods but is also the most computationally efficient by far and thus may be the only viable option for edge applications with severe hardware constraints. 

End-to-end approaches on the other hand offer better computational efficiency than all clustering approaches except the \textcolor{black}{AHC-BIC}. 
Overall EEND-VC performs the best among all methods and thus has the best trade-off between computing requirements and DER.
SA-EEND instead suffers from training versus inference sequence length mismatch, leading to performance degradation in recordings with longer duration. 
\textcolor{black}{Both SA-EEND and EEND-VC are highly susceptible to the characteristics of the training data and in particular to its size, and state-of-the-art results are only possible when trained on several thousands of hours of data.}

Finally, SSGD can achieve good performance but its performance degrades in presence of acoustic mismatches between training and test data. EEND-based approaches are instead more robust to this. 
\textcolor{black}{This is due to the fact that SSGD needs to be trained with datasets for which the oracle single speaker sources are available, a condition which is not usually satisfied on real-world data, leading to a possible domain mismatch. End-to-end fine-tuning could be a potential solution, and it could be explored in future works.}
Another drawback of SSGD is that it is the most demanding diarization technique computationally-wise among the ones considered here. This is because it relies on speech separation which is inherently a more challenging task than diarization. This aspect can only be tolerated if speech separation has another purpose aside from diarization (e.g., ASR).

\textcolor{black}{Our contribution cannot be considered exhaustive, like any experimental review, and our intention is that it can be found useful as a reference for future research directions and industrial applications. 
Possible future extensions could focus on different scenarios such as meeting scenarios where the overlap conditions and the number of speakers are different as well as include other diarization systems such as ones based on TS-VAD.}\\

\section{Acknowledgements}
\label{sec:ack}
We would like to thank Dr. Keisuke Kinoshita at NTT Communication Science Laboratories for providing the pre-trained EEND-VC model.

This work has been supported by the AGEVOLA project (SIME code 2019.0227), funded by the Fondazione CARITRO.

\bibliographystyle{elsarticle-num} 
\bibliography{main_clean.bib}

\begin{thebibliography}{100}
\expandafter\ifx\csname url\endcsname\relax
  \def\url#1{\texttt{#1}}\fi
\expandafter\ifx\csname urlprefix\endcsname\relax\def\urlprefix{URL }\fi
\expandafter\ifx\csname href\endcsname\relax
  \def\href#1#2{#2} \def\path#1{#1}\fi

\bibitem{boeddeker2018front}
C.~Boeddeker, J.~Heitkaemper, J.~Schmalenstroeer, L.~Drude, J.~Heymann,
  R.~Haeb-Umbach, {Front-end processing for the CHiME-5 dinner party scenario},
  in: Proc. of CHiME-5 Workshop on Speech Processing in Everyday Environments,
  2018, pp. 35--40.

\bibitem{kanda2019guided}
N.~Kanda, C.~Boeddeker, J.~Heitkaemper, Y.~Fujita, S.~Horiguchi,
  R.~Haeb-Umbach, Guided source separation meets a strong {ASR} backend:
  {Hitachi/Paderborn University} joint investigation for dinner party {ASR},
  in: Proc. of Interspeech, 2019, pp. 1248--1252.

\bibitem{saon2013speaker}
G.~Saon, H.~Soltau, D.~Nahamoo, M.~Picheny, Speaker adaptation of neural
  network acoustic models using i-vectors, in: Proc. of ASRU, IEEE, 2013, pp.
  55--59.

\bibitem{miao2015speaker}
Y.~Miao, H.~Zhang, F.~Metze, Speaker adaptive training of deep neural network
  acoustic models using i-vectors, IEEE/ACM Transactions on Audio, Speech, and
  Language Processing 23~(11) (2015) 1938--1949.

\bibitem{wang2017unsupervised}
Z.-Q. Wang, D.~Wang, Unsupervised speaker adaptation of batch normalized
  acoustic models for robust {ASR}, in: Proc. of ICASSP, IEEE, 2017, pp.
  4890--4894.

\bibitem{sari2020unsupervised}
L.~Sar{\i}, N.~Moritz, T.~Hori, J.~Le~Roux, Unsupervised speaker adaptation
  using attention-based speaker memory for end-to-end {ASR}, in: Proc. of
  ICASSP, IEEE, 2020, pp. 7384--7388.

\bibitem{fujita2019end-blstm}
Y.~Fujita, N.~Kanda, S.~Horiguchi, K.~Nagamatsu, S.~Watanabe, End-to-end neural
  speaker diarization with permutation-free objectives, in: Proc. of
  Interspeech, 2019, pp. 4300--4304.

\bibitem{fujita2019end-self}
Y.~Fujita, N.~Kanda, S.~Horiguchi, Y.~Xue, K.~Nagamatsu, S.~Watanabe,
  End-to-end neural speaker diarization with self-attention, in: Proc. of ASRU,
  IEEE, 2019, pp. 296--303.

\bibitem{fujita2020end}
Y.~Fujita, S.~Watanabe, S.~Horiguchi, Y.~Xue, K.~Nagamatsu, End-to-end neural
  diarization: Reformulating speaker diarization as simple multi-label
  classification, arXiv preprint arXiv:2003.02966 (2020).

\bibitem{kinoshita2021integrating}
K.~Kinoshita, M.~Delcroix, N.~Tawara, Integrating end-to-end neural and
  clustering-based diarization: Getting the best of both worlds, in: Proc. of
  ICASSP, IEEE, 2021, pp. 7198--7202.

\bibitem{kinoshita2021advances}
K.~Kinoshita, M.~Delcroix, N.~Tawara, Advances in integration of end-to-end
  neural and clustering-based diarization for real conversational speech, in:
  Proc. of Interspeech, 2021, pp. 3565--3569.

\bibitem{kinoshita2022utterance}
K.~Kinoshita, T.~von Neumann, M.~Delcroix, C.~Boeddeker, R.~Haeb-Umbach,
  {Utterance-by-utterance overlap-aware neural diarization with Graph-PIT}, in:
  Proc. of Interspeech, 2022, pp. 1486--1490.

\bibitem{kinoshita2022tight}
K.~Kinoshita, M.~Delcroix, T.~Iwata, Tight integration of neural-and
  clustering-based diarization through deep unfolding of infinite gaussian
  mixture model, in: Proc. of ICASSP, IEEE, 2022, pp. 8382--8386.

\bibitem{watanabe2020chime}
S.~Watanabe, M.~Mandel, J.~Barker, E.~Vincent, A.~Arora, X.~Chang,
  S.~Khudanpur, V.~Manohar, D.~Povey, D.~Raj, D.~Snyder, A.~S. Subramanian,
  J.~Trmal, B.~B. Yair, C.~Boeddeker, Z.~Ni, Y.~Fujita, S.~Horiguchi, N.~Kanda,
  T.~Yoshioka, N.~Ryant, {CHiME-6} challenge: Tackling multispeaker speech
  recognition for unsegmented recordings, in: Proc. of CHiME-6 Workshop on
  Speech Processing in Every- day Environments, 2020, pp. 1--7.

\bibitem{gish1991segregation}
H.~Gish, M.-H. Siu, J.~R. Rohlicek, Segregation of speakers for speech
  recognition and speaker identification, in: Proc. of ICASSP, 1991, pp.
  873--876.

\bibitem{siu1992unsupervised}
M.-H. Siu, G.~Yu, H.~Gish, An unsupervised, sequential learning algorithm for
  the segmentation of speech waveforms with multiple speakers, in: Proc. of
  ICASSP, IEEE, 1992, pp. 189--192.

\bibitem{jain1996recognition}
U.~Jain, M.~A. Siegler, S.-J. Doh, E.~Gouvea, J.~Huerta, P.~J. Moreno, B.~Raj,
  R.~M. Stern, Recognition of continuous broadcast news with multiple unknown
  speakers and environments, in: Proc. of DARPA Speech Recognition Workshop,
  1996, pp. 61--66.

\bibitem{chen1998speaker}
S.~Chen, P.~Gopalakrishnan, et~al., Speaker, environment and channel change
  detection and clustering via the bayesian information criterion, in: Proc. of
  DARPA broadcast news transcription and understanding workshop, 1998, pp.
  127--132\color{black}.

\bibitem{liu1999fast}
D.~Liu, F.~Kubala, Fast speaker change detection for broadcast news
  transcription and indexing, in: Proc. of Eurospeech, 1999, pp. 1031--1034.

\bibitem{hermansky1990perceptual}
H.~Hermansky, Perceptual linear predictive {(PLP)} analysis of speech, The
  Journal of the Acoustical Society of America 87~(4) (1990) 1738--1752.

\bibitem{o1988linear}
D.~O'Shaughnessy, Linear predictive coding, IEEE Potentials 7~(1) (1988)
  29--32.

\bibitem{yamaguchi2005spectral}
M.~Yamaguchi, M.~Yamashita, S.~Matsunaga, Spectral cross-correlation features
  for audio indexing of broadcast news and meetings, in: Proc. of Interspeech,
  2005, pp. 613--616.

\bibitem{carletta2005ami}
J.~Carletta, S.~Ashby, S.~Bourban, M.~Flynn, M.~Guillemot, T.~Hain, J.~Kadlec,
  V.~Karaiskos, W.~Kraaij, M.~Kronenthal, et~al., The {AMI} meeting corpus: A
  pre-announcement, in: International workshop on machine learning for
  multimodal interaction, Springer, 2005, pp. 28--39.

\bibitem{reynolds2000speaker}
D.~A. Reynolds, T.~F. Quatieri, R.~B. Dunn, Speaker verification using adapted
  gaussian mixture models, Digital Signal Processing 10~(1-3) (2000) 19--41.

\bibitem{kenny2007speaker}
P.~Kenny, G.~Boulianne, P.~Ouellet, P.~Dumouchel, Speaker and session
  variability in {GMM}-based speaker verification, IEEE/ACM Transactions on
  Audio, Speech, and Language Processing 15~(4) (2007) 1448--1460.

\bibitem{castaldo2008stream}
F.~Castaldo, D.~Colibro, E.~Dalmasso, P.~Laface, C.~Vair, Stream-based speaker
  segmentation using speaker factors and eigenvoices, in: Proc. of ICASSP,
  IEEE, 2008, pp. 4133--4136.

\bibitem{dehak2010front}
N.~Dehak, P.~J. Kenny, R.~Dehak, P.~Dumouchel, P.~Ouellet, Front-end factor
  analysis for speaker verification, IEEE/ACM Transactions on Audio, Speech,
  and Language Processing 19~(4) (2010) 788--798.

\bibitem{hatch2006within}
A.~O. Hatch, S.~Kajarekar, A.~Stolcke, Within-class covariance normalization
  for {SVM}-based speaker recognition, in: Proc. of Interspeech, 2006, pp.
  1471--1474.

\bibitem{matvejka2011full}
P.~Mat{\v{e}}jka, O.~Glembek, F.~Castaldo, M.~J. Alam, O.~Plchot, P.~Kenny,
  L.~Burget, J.~{\v{C}}ernocky, Full-covariance {UBM} and heavy-tailed {PLDA}
  in i-vector speaker verification, in: Proc. of ICASSP, IEEE, 2011, pp.
  4828--4831.

\bibitem{variani2014deep}
E.~Variani, X.~Lei, E.~McDermott, I.~L. Moreno, J.~Gonzalez-Dominguez, Deep
  neural networks for small footprint text-dependent speaker verification, in:
  Proc. of ICASSP, IEEE, 2014, pp. 4052--4056.

\bibitem{bai2021speaker}
Z.~Bai, X.-L. Zhang, Speaker recognition based on deep learning: An overview,
  Neural Networks 140 (2021) 65--99.

\bibitem{snyder2018x}
D.~Snyder, D.~Garcia-Romero, G.~Sell, D.~Povey, S.~Khudanpur, X-vectors: Robust
  {DNN} embeddings for speaker recognition, in: Proc. of ICASSP, IEEE, 2018,
  pp. 5329--5333.

\bibitem{landini2021analysis}
F.~Landini, O.~Glembek, P.~Mat{\v{e}}jka, J.~Rohdin, L.~Burget, M.~Diez,
  A.~Silnova, Analysis of the {BUT} diarization system for {VoxConverse}
  challenge, in: Proc. of ICASSP, IEEE, 2021, pp. 5819--5823.

\bibitem{koluguri2022titanet}
N.~R. Koluguri, T.~Park, B.~Ginsburg, {TitaNet}: Neural model for speaker
  representation with 1d depth-wise separable convolutions and global context,
  in: Proc. of ICASSP, IEEE, 2022, pp. 8102--8106.

\bibitem{desplanques2020ecapa}
B.~Desplanques, J.~Thienpondt, K.~Demuynck, {ECAPA-TDNN}: Emphasized channel
  attention, propagation and aggregation in {TDNN} based speaker verification,
  in: Proc. of Interspeech, 2020, pp. 3830--3834.

\bibitem{chen2022wavlm}
S.~Chen, C.~Wang, Z.~Chen, Y.~Wu, S.~Liu, Z.~Chen, J.~Li, N.~Kanda,
  T.~Yoshioka, X.~Xiao, et~al., {WavLM}: Large-scale self-supervised
  pre-training for full stack speech processing, IEEE Journal of Selected
  Topics in Signal Processing 16~(6) (2022) 1505--1518.

\bibitem{bredin2017tristounet}
H.~Bredin, {TristouNet}: triplet loss for speaker turn embedding, in: Proc. of
  ICASSP, IEEE, 2017, pp. 5430--5434.

\bibitem{li2018angular}
Y.~Li, F.~Gao, Z.~Ou, J.~Sun, Angular softmax loss for end-to-end speaker
  verification, in: Proc. of ISCSLP, IEEE, 2018, pp. 190--194.

\bibitem{chung2020defence}
J.~S. Chung, J.~Huh, S.~Mun, M.~Lee, H.~S. Heo, S.~Choe, C.~Ham, S.~Jung, B.-J.
  Lee, I.~Han, In defence of metric learning for speaker recognition, in: Proc.
  of Interspeech, 2020, pp. 2977--2981.

\bibitem{ng2001spectral}
A.~Ng, M.~Jordan, Y.~Weiss, On spectral clustering: Analysis and an algorithm,
  Advances in neural information processing systems 14 (2001).

\bibitem{ning2006spectral}
H.~Ning, M.~Liu, H.~Tang, T.~S. Huang, A spectral clustering approach to
  speaker diarization, in: Ninth International Conference on Spoken Language
  Processing, 2006.

\bibitem{wang2018speaker}
Q.~Wang, C.~Downey, L.~Wan, P.~A. Mansfield, I.~L. Moreno, Speaker diarization
  with lstm, in: 2018 IEEE International conference on acoustics, speech and
  signal processing (ICASSP), IEEE, 2018, pp. 5239--5243\color{black}.

\bibitem{park2019auto}
T.~J. Park, K.~J. Han, M.~Kumar, S.~Narayanan, Auto-tuning spectral clustering
  for speaker diarization using normalized maximum eigengap, IEEE Signal
  Processing Letters 27 (2019) 381--385.

\bibitem{diez2018speaker}
M.~Diez, L.~Burget, P.~Matejka, Speaker diarization based on bayesian {HMM}
  with eigenvoice priors, in: Proc. of Odyssey, 2018, pp. 147--154.

\bibitem{landini2020bayesian}
F.~Landini, J.~Profant, M.~Diez, L.~Burget, Bayesian {HMM} clustering of
  x-vector sequences {(VBx)} in speaker diarization: theory, implementation and
  analysis on standard tasks, Computer Speech \& Language 71 (2020) 101254.

\bibitem{landini2022simulated}
F.~Landini, A.~Lozano-Diez, M.~Diez, L.~Burget, From simulated mixtures to
  simulated conversations as training data for end-to-end neural diarization,
  2022, pp. 5095--5099.

\bibitem{ryant2019second}
N.~Ryant, K.~Church, C.~Cieri, A.~Cristia, J.~Du, S.~Ganapathy, M.~Liberman,
  Second {DIHARD} challenge evaluation plan, Linguistic Data Consortium, Tech.
  Rep (2019).

\bibitem{stolcke2019dover}
A.~Stolcke, T.~Yoshioka, {DOVER}: A method for combining diarization outputs,
  in: Proc. of ASRU, IEEE, 2019, pp. 757--763.

\bibitem{raj2021dover}
D.~Raj, L.~P. Garcia-Perera, Z.~Huang, S.~Watanabe, D.~Povey, A.~Stolcke,
  S.~Khudanpur, {DOVER-Lap}: A method for combining overlap-aware diarization
  outputs, in: Proc. of SLT, IEEE, 2021, pp. 881--888.

\bibitem{raj2021reformulating}
D.~Raj, S.~Khudanpur, Reformulating {DOVER-Lap} label mapping as a graph
  partitioning problem, in: Proc. of Interspeech, 2021, pp. 2351--2355.

\bibitem{bullock2020overlap}
L.~Bullock, H.~Bredin, L.~P. Garcia-Perera, Overlap-aware diarization:
  Resegmentation using neural end-to-end overlapped speech detection, in: Proc.
  of ICASSP, IEEE, 2020, pp. 7114--7118.

\bibitem{raj2021multi}
D.~Raj, Z.~Huang, S.~Khudanpur, Multi-class spectral clustering with overlaps
  for speaker diarization, in: Proc. of SLT, IEEE, 2021, pp. 582--589.

\bibitem{huang2020speaker}
Z.~Huang, S.~Watanabe, Y.~Fujita, P.~Garc{\'\i}a, Y.~Shao, D.~Povey,
  S.~Khudanpur, Speaker diarization with region proposal network, in: Proc. of
  ICASSP, IEEE, 2020, pp. 6514--6518.

\bibitem{zhang2019fully}
A.~Zhang, Q.~Wang, Z.~Zhu, J.~Paisley, C.~Wang, Fully supervised speaker
  diarization, in: Proc. of ICASSP, IEEE, 2019, pp. 6301--6305.

\bibitem{li2021discriminative}
Q.~Li, F.~L. Kreyssig, C.~Zhang, P.~C. Woodland, Discriminative neural
  clustering for speaker diarisation, in: 2021 IEEE Spoken Language Technology
  Workshop (SLT), IEEE, 2021, pp. 574--581\color{black}.

\bibitem{fang2021deep}
X.~Fang, Z.-H. Ling, L.~Sun, S.-T. Niu, J.~Du, C.~Liu, Z.-C. Sheng, A deep
  analysis of speech separation guided diarization under realistic conditions,
  in: Proc. of APSIPA ASC, IEEE, 2021, pp. 667--671.

\bibitem{morrone2022low-latency}
G.~Morrone, S.~Cornell, D.~Raj, L.~Serafini, E.~Zovato, A.~Brutti,
  S.~Squartini, Low-latency speech separation guided diarization for telephone
  conversations, in: Proc. of SLT, IEEE, 2022.

\bibitem{medennikov2020target}
I.~Medennikov, M.~Korenevsky, T.~Prisyach, Y.~Khokhlov, M.~Korenevskaya,
  I.~Sorokin, T.~Timofeeva, A.~Mitrofanov, A.~Andrusenko, I.~Podluzhny, et~al.,
  Target-speaker voice activity detection: A novel approach for multi-speaker
  diarization in a dinner party scenario, in: Proc. of Interspeech, 2020, pp.
  274--278.

\bibitem{ryant2020third}
N.~Ryant, K.~Church, C.~Cieri, J.~Du, S.~Ganapathy, M.~Liberman, Third {DIHARD}
  challenge evaluation plan, in: Proc. of Interspeech, 2021, pp. 3570--3574.

\bibitem{wang2021ustc}
Y.~Wang, M.~He, S.~Niu, L.~Sun, T.~Gao, X.~Fang, J.~Pan, J.~Du, C.-H. Lee,
  {USTC-NELSLIP} system description for {DIHARD-III} challenge, in: Proc. of
  3rd DIHARD Speech Diarization Challenge Workshop, 2021\color{black}.

\bibitem{brown2022voxsrc}
A.~Brown, J.~Huh, J.~S. Chung, A.~Nagrani, D.~Garcia-Romero, A.~Zisserman,
  Voxsrc 2021: The third voxceleb speaker recognition challenge, arXiv preprint
  arXiv:2201.04583 (2022).

\bibitem{huh2023voxsrc}
J.~Huh, A.~Brown, J.-w. Jung, J.~S. Chung, A.~Nagrani, D.~Garcia-Romero,
  A.~Zisserman, Voxsrc 2022: The fourth voxceleb speaker recognition challenge,
  arXiv preprint arXiv:2302.10248 (2023).

\bibitem{yu2022m2met}
F.~Yu, S.~Zhang, Y.~Fu, L.~Xie, S.~Zheng, Z.~Du, W.~Huang, P.~Guo, Z.~Yan,
  B.~Ma, X.~Xu, H.~Bu, M2met: The icassp 2022 multi-channel multi-party meeting
  transcription challenge, in: ICASSP 2022 - 2022 IEEE International Conference
  on Acoustics, Speech and Signal Processing (ICASSP), 2022, pp. 6167--6171.
\newblock \href {https://doi.org/10.1109/ICASSP43922.2022.9746465}
  {\path{doi:10.1109/ICASSP43922.2022.9746465}}.

\bibitem{wang2021dku}
W.~Wang, D.~Cai, Q.~Lin, L.~Yang, J.~Wang, J.~Wang, M.~Li, The
  dku-dukeece-lenovo system for the diarization task of the 2021 voxceleb
  speaker recognition challenge, arXiv preprint arXiv:2109.02002 (2021).

\bibitem{wang2022dku}
W.~Wang, X.~Qin, M.~Cheng, Y.~Zhang, K.~Wang, M.~Li, The dku-dukeece
  diarization system for the voxceleb speaker recognition challenge 2022, arXiv
  preprint arXiv:2210.01677 (2022).

\bibitem{wang2022similarity}
W.~Wang, Q.~Lin, D.~Cai, M.~Li, Similarity measurement of segment-level speaker
  embeddings in speaker diarization, IEEE/ACM Transactions on Audio, Speech,
  and Language Processing 30 (2022) 2645--2658.

\bibitem{wang2022cross}
W.~Wang, X.~Qin, M.~Li, Cross-channel attention-based target speaker voice
  activity detection: Experimental results for the m2met challenge, in: Proc.
  of ICASSP, IEEE, 2022, pp. 9171--9175.

\bibitem{wang2022incorporating}
W.~Wang, M.~Li, Incorporating end-to-end framework into target-speaker voice
  activity detection, in: Proc. of ICASSP, IEEE, 2022, pp. 8362--8366.

\bibitem{wang2022online}
W.~Wang, Q.~Lin, M.~Li, Online target speaker voice activity detection for
  speaker diarization, arXiv preprint arXiv:2207.05920 (2022).

\bibitem{cheng2022target}
M.~Cheng, W.~Wang, Y.~Zhang, X.~Qin, M.~Li, Target-speaker voice activity
  detection via sequence-to-sequence prediction, arXiv preprint
  arXiv:2210.16127 (2022\color{black}).

\bibitem{kolbaek2017multitalker}
M.~Kolb{\ae}k, D.~Yu, Z.-H. Tan, J.~Jensen, Multitalker speech separation with
  utterance-level permutation invariant training of deep recurrent neural
  networks, IEEE/ACM Transactions on Audio, Speech, and Language Processing
  25~(10) (2017) 1901--1913.

\bibitem{horiguchi2020end}
S.~Horiguchi, Y.~Fujita, S.~Watanabe, Y.~Xue, K.~Nagamatsu, End-to-end speaker
  diarization for an unknown number of speakers with encoder-decoder based
  attractors, in: Proc. of Interspeech, 2020, pp. 269--273.

\bibitem{horiguchi2022encoder}
S.~Horiguchi, Y.~Fujita, S.~Watanabe, Y.~Xue, P.~Garcia, Encoder-decoder based
  attractors for end-to-end neural diarization, IEEE/ACM Transactions on Audio,
  Speech, and Language Processing 30 (2022) 1493--1507\color{black}.

\bibitem{xue2021online}
Y.~Xue, S.~Horiguchi, Y.~Fujita, S.~Watanabe, P.~Garc{\'\i}a, K.~Nagamatsu,
  Online end-to-end neural diarization with speaker-tracing buffer, in: Proc.
  of SLT, IEEE, 2021, pp. 841--848.

\bibitem{han2021bw}
E.~Han, C.~Lee, A.~Stolcke, {BW-EDA-EEND}: Streaming end-to-end neural speaker
  diarization for a variable number of speakers, in: Proc. of ICASSP, IEEE,
  2021, pp. 7193--7197.

\bibitem{horiguchi2022online}
S.~Horiguchi, S.~Watanabe, P.~Garcia, Y.~Takashima, Y.~Kawaguchi, Online neural
  diarization of unlimited numbers of speakers, arXiv preprint arXiv:2206.02432
  (2022).

\bibitem{coria2021overlap}
J.~M. Coria, H.~Bredin, S.~Ghannay, S.~Rosset, Overlap-aware low-latency online
  speaker diarization based on end-to-end local segmentation, in: Proc. of
  ASRU, IEEE, 2021, pp. 1139--1146.

\bibitem{horiguchi2021hitachi}
S.~Horiguchi, N.~Yalta, P.~Garcia, Y.~Takashima, Y.~Xue, D.~Raj, Z.~Huang,
  Y.~Fujita, S.~Watanabe, S.~Khudanpur, {The Hitachi-JHU DIHARD III system:
  Competitive end-to-end neural diarization and x-vector clustering systems
  combined by DOVER-Lap}, in: 3rd DIHARD Speech Diarization Challenge Workshop,
  2021.

\bibitem{landini2021but}
F.~Landini, A.~Lozano-Diez, L.~Burget, M.~Diez, A.~Silnova,
  K.~Zmol{\i}kov{\'a}, O.~Glembek, P.~Matejka, T.~Stafylakis, N.~Br{\"u}mmer,
  {BUT system description for the third DIHARD speech diarization challenge},
  in: Proc. of 3rd DIHARD Speech Diarization Challenge Workshop, 2021.

\bibitem{zeghidour2021dive}
N.~Zeghidour, O.~Teboul, D.~Grangier, {DIVE}: End-to-end speech diarization via
  iterative speaker embedding, in: Proc. of ASRU, IEEE, 2021, pp. 702--709.

\bibitem{shafey2019joint}
L.~E. Shafey, H.~Soltau, I.~Shafran, {Joint Speech Recognition and Speaker
  Diarization via Sequence Transduction}, in: Proc. Interspeech 2019, 2019, pp.
  396--400\color{black}.

\bibitem{tranter2006overview}
S.~E. Tranter, D.~A. Reynolds, An overview of automatic speaker diarization
  systems, IEEE/ACM Transactions on Audio, Speech, and Language Processing
  14~(5) (2006) 1557--1565.

\bibitem{anguera2012speaker}
X.~Anguera, S.~Bozonnet, N.~Evans, C.~Fredouille, G.~Friedland, O.~Vinyals,
  Speaker diarization: A review of recent research, IEEE/ACM Transactions on
  Audio, Speech, and Language Processing 20~(2) (2012) 356--370.

\bibitem{park2022review}
T.~J. Park, N.~Kanda, D.~Dimitriadis, K.~J. Han, S.~Watanabe, S.~Narayanan, A
  review of speaker diarization: Recent advances with deep learning, Computer
  Speech \& Language 72 (2022) 101317.

\bibitem{schwartz2020green}
R.~Schwartz, J.~Dodge, N.~A. Smith, O.~Etzioni, Green {AI}, Communications of
  the ACM 63~(12) (2020) 54--63.

\bibitem{diez2019analysis}
M.~Diez, L.~Burget, F.~Landini, J.~{\v{C}}ernock{\`y}, Analysis of speaker
  diarization based on bayesian {HMM} with eigenvoice priors, IEEE/ACM
  Transactions on Audio, Speech, and Language Processing 28 (2019) 355--368.

\bibitem{lloyd1982least}
S.~Lloyd, Least squares quantization in {PCM}, IEEE/ACM Transactions on Audio,
  Speech, and Language Processing 28~(2) (1982) 129--137.

\bibitem{tritschler1999improved}
A.~Tritschler, R.~A. Gopinath, Improved speaker segmentation and segments
  clustering using the bayesian information criterion, in: Proc. of Eurospeech,
  1999, pp. 679--682.

\bibitem{xia2022turn}
W.~Xia, H.~Lu, Q.~Wang, A.~Tripathi, Y.~Huang, I.~L. Moreno, H.~Sak,
  Turn-to-diarize: Online speaker diarization constrained by transformer
  transducer speaker turn detection, in: ICASSP 2022-2022 IEEE International
  Conference on Acoustics, Speech and Signal Processing (ICASSP), IEEE, 2022,
  pp. 8077--8081\color{black}.

\bibitem{chen2020continuous}
Z.~Chen, T.~Yoshioka, L.~Lu, T.~Zhou, Z.~Meng, Y.~Luo, J.~Wu, X.~Xiao, J.~Li,
  Continuous speech separation: Dataset and analysis, in: Proc. of ICASSP,
  IEEE, 2020, pp. 7284--7288.

\bibitem{wagstaff2001constrained}
K.~Wagstaff, C.~Cardie, S.~Rogers, S.~Schr{\"o}dl, et~al., Constrained
  {K}-means clustering with background knowledge, in: Proc. of ICML, 2001, pp.
  577--584.

\bibitem{zeghidour2021wavesplit}
N.~Zeghidour, D.~Grangier, Wavesplit: End-to-end speech separation by speaker
  clustering, IEEE/ACM Transactions on Audio, Speech, and Language Processing
  29 (2021) 2840--2849.

\bibitem{callhome}
M.~Przybocki, M.~Alvin, \href{https://catalog.ldc.upenn.edu/LDC2001S97}{2000
  {NIST} {Speaker} {Recognition} {Evaluation} {LDC2001S9}} (2001).
\newline\urlprefix\url{https://catalog.ldc.upenn.edu/LDC2001S97}

\bibitem{cieri2004fisher}
C.~Cieri, D.~Miller, K.~Walker, The {Fisher} corpus: A resource for the next
  generations of speech-to-text, in: LREC, Vol.~4, 2004, pp. 69--71.

\bibitem{reddy2021dnsmos}
C.~K. Reddy, V.~Gopal, R.~Cutler, {DNSMOS}: A non-intrusive perceptual
  objective speech quality metric to evaluate noise suppressors, in: Proc. of
  ICASSP, IEEE, 2021, pp. 6493--6497.

\bibitem{peddinti2015jhu}
V.~Peddinti, G.~Chen, V.~Manohar, T.~Ko, D.~Povey, S.~Khudanpur, {JHU ASpIRE
  system: Robust LVCSR with TDNNs, iVector adaptation and RNN-LMS}, in: Proc.
  of ASRU, IEEE, 2015, pp. 539--546.

\bibitem{he2016deep}
K.~He, X.~Zhang, S.~Ren, J.~Sun, Deep residual learning for image recognition,
  in: Proc. of CVPR, 2016, pp. 770--778.

\bibitem{nagrani2017voxceleb}
A.~Nagrani, J.~S. Chung, A.~Zisserman, {VoxCeleb}: A large-scale speaker
  identification dataset, in: Proc. of Interspeech, 2017, pp. 2616--2620.

\bibitem{chung2018voxceleb2}
J.~S. Chung, A.~Nagrani, A.~Zisserman, {VoxCeleb2}: Deep speaker recognition,
  in: Proc. of Interspeech, 2018, pp. 1086--1090.

\bibitem{fan2020cn}
Y.~Fan, J.~Kang, L.~Li, K.~Li, H.~Chen, S.~Cheng, P.~Zhang, Z.~Zhou, Y.~Cai,
  D.~Wang, {CN-Celeb}: a challenging chinese speaker recognition dataset, in:
  Proc. of ICASSP, IEEE, 2020, pp. 7604--7608.

\bibitem{zeinali2018deepmine}
H.~Zeinali, H.~Sameti, T.~Stafylakis, {DeepMine} speech processing database:
  Text-dependent and independent speaker verification and speech recognition in
  {Persian} and {English}, in: Proc. of Odyssey, 2018, pp. 386--392.

\bibitem{snyder2015musan}
D.~Snyder, G.~Chen, D.~Povey, {MUSAN}: A music, speech, and noise corpus, arXiv
  preprint arXiv:1510.08484 (2015).

\bibitem{bredin2021end}
H.~Bredin, A.~Laurent, End-to-end speaker segmentation for overlap-aware
  resegmentation, in: Proc. of Interspeech, 2021, pp. 3111--3115.

\bibitem{bredin2020pyannote}
H.~Bredin, R.~Yin, J.~M. Coria, G.~Gelly, P.~Korshunov, M.~Lavechin, D.~Fustes,
  H.~Titeux, W.~Bouaziz, M.-P. Gill, {Pyannote.Audio}: neural building blocks
  for speaker diarization, in: Proc. of ICASSP, IEEE, 2020, pp. 7124--7128.

\bibitem{junqua1994robust}
J.-C. Junqua, B.~Mak, B.~Reaves, A robust algorithm for word boundary detection
  in the presence of noise, IEEE Transactions on Speech and Audio Processing
  2~(3) (1994) 406--412.

\bibitem{luo2020dual}
Y.~Luo, Z.~Chen, T.~Yoshioka, {Dual-Path RNN}: efficient long sequence modeling
  for time-domain single-channel speech separation, in: Proc. of ICASSP, IEEE,
  2020, pp. 46--50.

\bibitem{cornell2022overlapped}
S.~Cornell, M.~Omologo, S.~Squartini, E.~Vincent, Overlapped speech detection
  and speaker counting using distant microphone arrays, Computer Speech \&
  Language 72 (2022) 101306.

\bibitem{luo2019conv}
Y.~Luo, N.~Mesgarani, {Conv-TasNet}: Surpassing ideal time--frequency magnitude
  masking for speech separation, IEEE/ACM Transactions on Audio, Speech, and
  Language Processing 27~(8) (2019) 1256--1266.

\bibitem{paszke2019pytorch}
A.~Paszke, S.~Gross, F.~Massa, A.~Lerer, J.~Bradbury, G.~Chanan, T.~Killeen,
  Z.~Lin, N.~Gimelshein, L.~Antiga, et~al., Pytorch: An imperative style,
  high-performance deep learning library, Advances in NIPS 32 (2019)
  8026–--8037.

\bibitem{kingma2015adam}
D.~P. Kingma, J.~Ba, Adam: A method for stochastic optimization, in: Proc. of
  ICLR, 2015.

\bibitem{vaswani2017attention}
A.~Vaswani, N.~Shazeer, N.~Parmar, J.~Uszkoreit, L.~Jones, A.~N. Gomez,
  {\L}.~Kaiser, I.~Polosukhin, Attention is all you need, Advances in NIPS 30
  (2017) 6000–--6010.

\end{thebibliography}
\end{document}